\documentclass[twocolumn]{aastex7}

\usepackage{enumitem}
\usepackage{amsmath}

\begin{document}

\title{SCALES. II. The Internal Composition of Giant Star-Forming Clumps at $z = 1 - 4$}

\correspondingauthor{Boris S. Kalita}
\email{boris.kalita@pku.edu.cn; kalita.boris.sindhu@gmail.com}

\author[0000-0001-9215-7053]{Boris S. Kalita}
\altaffiliation{Kavli Astrophysics Fellow and Special Boya Fellow}
\affiliation{Kavli Institute for Astronomy and Astrophysics, Peking University, Beijing 100871, People's Republic of China}
\affiliation{Kavli IPMU (WPI), UTIAS, The University of Tokyo, Kashiwa, Chiba 277-8583, Japan}
\email{boris.kalita@pku.edu.cn; kalita.boris.sindhu@gmail.com}

\author[0000-0001-7940-1816]{Adélaïde Claeyssens}
\affiliation{Univ Lyon, Univ Lyon1, ENS de Lyon, CNRS, Centre de Recherche Astrophysique de Lyon UMR5574, F-69230 Saint-Genis-Laval, France}
\email{adelaide.claeyssens@univ-lyon1.fr}

\author[0000-0001-6947-5846]{Luis C. Ho}
\affiliation{Kavli Institute for Astronomy and Astrophysics, Peking University, Beijing 100871, People's Republic of China}
\affiliation{Department of Astronomy, School of Physics, Peking University, Beijing 100871, People's Republic of China}
\email{lho.pku@gmail.com}

\author[0000-0003-0348-2917]{Miroslava Dessauges-Zavadsky}
\affiliation{Département d'Astronomie, Université de Genève, Chemin Pegasi 51, CH-1290 Versoix, Switzerland}
\email{Miroslava.Dessauges@unige.ch}

\author[0000-0003-4264-3381]{Natascha M. Förster Schreiber}
\affiliation{Max-Planck-Institut fur extraterrestrische Physik, Giessenbachstrasse, D-85748 Garching, Germany}
\email{forster@mpe.mpg.de}

\author[0000-0002-3331-9590]{Emanuele Daddi}
\affiliation{Universit\'e Paris-Saclay, Universit\'e Paris Cit\'e, CEA, CNRS, AIM, 91191, Gif-sur-Yvette, France}
\email{emanuele.daddi@cea.fr}

\author[0000-0002-8434-880X]{John D. Silverman}
\affiliation{Kavli IPMU (WPI), UTIAS, The University of Tokyo, Kashiwa, Chiba 277-8583, Japan}
\affiliation{Department of Astronomy, School of Science, The University of Tokyo, 7-3-1 Hongo, Bunkyo, Tokyo 113-0033, Japan}
\affiliation{Centre for Data-Driven Discovery, Kavli IPMU (WPI), UTIAS, The University of Tokyo, Kashiwa, Chiba 277-8583, Japan}
\affiliation{Centre for Astrophysical Sciences, Department of Physics \& Astronomy, Johns Hopkins University, Baltimore, MD 21218, USA}
\email{john.silverman@ipmu.jp}

\author[0000-0001-8812-8460]{Naval Kishor Bhadari}
\affiliation{Kavli Institute for Astronomy and Astrophysics, Peking University, Beijing 100871, People's Republic of China}
\email{naval@pku.edu.cn}

\author[0009-0003-4721-177X]{Changhao Chen}
\affiliation{Kavli Institute for Astronomy and Astrophysics, Peking University, Beijing 100871, People's Republic of China}
\affiliation{Department of Astronomy, School of Physics, Peking University, Beijing 100871, People's Republic of China}
\email{cchen_louis@stu.pku.edu.cn}

\author[0000-0001-6115-0633]{Fangzhou Jiang}
\affiliation{Kavli Institute for Astronomy and Astrophysics, Peking University, Beijing 100871, People's Republic of China}
\affiliation{Department of Astronomy, School of Physics, Peking University, Beijing 100871, People's Republic of China}
\email{fangzhou.jiang@pku.edu.cn}

\author[0000-0002-7237-3856]{Ke Wang}
\affiliation{Kavli Institute for Astronomy and Astrophysics, Peking University, Beijing 100871, People's Republic of China}
\affiliation{Department of Astronomy, School of Physics, Peking University, Beijing 100871, People's Republic of China}
\email{kwang.astro@pku.edu.cn}

\author[0000-0001-8405-2921]{Jinning Liang}
\affiliation{Kavli Institute for Astronomy and Astrophysics, Peking University, Beijing 100871, People's Republic of China}
\affiliation{Department of Astronomy, School of Physics, Peking University, Beijing 100871, People's Republic of China}
\email{jinningliangastro@gmail.com}

\author[0000-0002-4639-5285]{Alvaro Segovia Otero}
\affiliation{Department of Astronomy, Tsinghua University, Beijing 100084, People's Republic of China}
\email{asegoviao@tsinghua.edu.cn}

\author[0000-0002-4569-9009]{Jinyi Shangguan}
\affiliation{Kavli Institute for Astronomy and Astrophysics, Peking University, Beijing 100871, People's Republic of China}
\affiliation{Department of Astronomy, School of Physics, Peking University, Beijing 100871, People's Republic of China}
\email{shangguan@pku.edu.cn}

\begin{abstract}

Star-forming structures of scale $\sim0.5-1\,\mathrm{kpc}$, dubbed `giant' clumps, are routinely detected in unlensed surveys of $z > 1$ galaxies. Explaining them requires characterizing their internal structure. In this work, part of the \emph{Star-Forming Clumps As Layered Emergent Structures} (\emph{SCALES}) series, we hence ask: \emph{What is inside a giant clump?} We analyze 23 lensed galaxies at $z=1-4$ with JWST/NIRCam using a difference-of-Gaussian decomposition to model (lensed) clumps at $\approx 50-400\,\mathrm{pc}$ scales, and the surrounding extended component (EC) at giant clump scales. Associating clumps to ECs, we identify 72 giant clumps. We find the constituent clumps above our mass completeness limit ($10^{7.1}\,\mathrm{M_\odot}$) in a giant clump to be consistent with random sampling of the clump stellar mass function (CSMF), hence holding no information about their host. This agrees qualitatively with both prevailing scenarios for giant clumps -- (I) long-lived structures with constantly renewed constituents, and (II) blended random clump associations. As most models lack predictions for their internal mass distribution, our observations provide an empirical constraint. Finally, the CSMF follows a power law of slope $1.88\pm0.04$ above a break at $\sim10^{7.5}\,\mathrm{M_\odot}$, close to our completeness limit. The giant clumps follow a consistent slope ($1.95\pm0.15$), giving the same scale-free mass function slope at (lensed) clump and giant clump scales. Within giant clumps, both the EC mass, if made up of undetected clumps, and the mass share of the most massive clump are consistent with the CSMF slope continuing down to $\approx10^{5.5}$--$10^{6}\,\mathrm{M_\odot}$, while the latter rejects an extension below $10^{5}\,\mathrm{M_\odot}$.  

\end{abstract}

\keywords{\uat{Galaxies}{573} --- }

\section{Introduction} \label{sec:intro}

In the local Universe, one finds the star-forming structures to be hierarchical and self-similar up to scales of hundreds of parsecs or even $\sim \mathrm{kpc}$ \citep{efremov95, larson95, elmegreen96a, elmegreen96b, elmegreen97,elmegreen06, bastian07, rodriguez20, larson23}. However, the largest scales are expected to be coherent for $< 100\,\mathrm{Myr}$, evidenced by a lack of spatial coherence over these scales when one isolates structures with older stellar populations \citep{gouliermis15, gouliermis17, grasha17, menon21,turner22,shashank25}. 

Meanwhile, the discussion on the hierarchical structure of star formation in the $z>1$ Universe is yet to be as rigorous, given the limited resolution available. Hence, studies usually focus on two disjoint spatial regimes.  At $\gtrsim500\,\mathrm{pc} $ to a few kpc, one finds giant star-forming `clumps'. These dominant structures \citep[initially observed using \textit{HST};][]{cowie95,elmegreen04,elmegreen07,elmegreen09} have been extensively studied for over two decades \citep{guo12,guo15,shibuya16,kalita25a,kalita25b,sok25,zhu26}. Their sizes are consistent with the largest coherent (Toomre) scales possible in the gas-rich host disks \citep{genzel08,genzel11,forsterschreiber09,wisnioski12,mieda16}. On the other hand,  studies targeting galaxies that have been gravitationally lensed rather identify clumps that are over an order of magnitude smaller \citep[$\sim 10-300\,\mathrm{pc}$;][]{jones10,livermore12,livermore15,dessaugeszavadsky17,vanzella21,mowla22,claeyssens23,adamo24,mowla24,messa24,claeyssens25,claeyssens26}, with those at $\ll100\,\mathrm{pc}$ requiring very high magnifications ($\mu\gg10$).

Using our understanding of local star-forming regions, it would be straightforward to assume that these two sets of observational studies could be the manifestation of some form of a star-forming hierarchy at $z>1$, spanning two separate regimes.  However, one cannot simply borrow the framework from the local Universe since star formation is occurring in extremely different interstellar medium conditions at these epochs \citep{genzel10,burkert10,carilli13, madau14,forsterschreiber20}, with higher gas fractions \citep{daddi10,tacconi10,magdis12, tacconi13,dessaugeszavadsky15,scoville16,scoville17,dessaugeszavadsky20} and gas turbulence \citep[][]{wisnioski15,ubler19, girard21, rizzo24,danhaive25}. Hence, an independent characterization of this star-forming hierarchy across different scales is required.

Understanding this hierarchy at $z>1$ goes beyond a fundamental understanding of star formation; it also potentially influences the morphological evolution of galaxies at $z>1$. If the structures detected at the largest scales, the `giant clumps', are truly the largest rung of this hierarchical ladder, and one accepts them to form out of a largely coherent collapse, their
$\sim10^{8-9}\,\mathrm{M_\odot}$ stellar masses \citep{guo15, kalita25a, kalita25b, sok25} could plausibly render them more bound and longer-lived (potentially surviving a few hundred Myr) than their local counterparts at similar scales, although whether such structures survive stellar feedback long enough to do so also remains debated \citep{genel12, mandelker14, mandelker17, oklopcic17, fensch21}.

Were a sufficient fraction to survive, some studies have claimed that it would lead to their inward migration due to dynamical friction, building the central mass concentration of the host galaxy while thickening the disk \citep{noguchi99, immeli04, bournaud07, dekel09, ceverino10, mandelker14, mandelker17, fensch21, dekel22}.  Meanwhile, the structures detected at smaller scales in lensed galaxies, the `clumps',  can be explained away as part of the cascading fragmentation within the same hierarchical framework \citep{faure21,genzel23}. Contrary to this scenario is the possibility that a giant clump is not a single coherent object but a physical aggregate of  the $\approx 50-400\,\mathrm{pc}$ clumps \citep{behrendt16, behrendt19}, or in the more extreme case a transient, largely blended chance superposition of randomly distributed clumps with little dynamical coherence \citep{tamburello15, tamburello17}.

The main challenge has been to bridge the two disjoint sets of observations -- those detecting giant clumps and those resolving lensed clumps. This is especially critical now that the most recent unlensed \emph{JWST} observations are reaching down toward lensed-clump scales. \cite{kalita25b}, for instance, measure clump sizes (effective radius) down to a few hundred parsecs, and even find indications of smaller structures embedded within the primary clumps, although the majority of their sample still lies in the giant clump regime. \cite{ji24} also find clump sizes (termed `substructures') down to similar values at redshifts $\sim 3-4$. These observations are hence a first step toward bridging the gap, since they raise the possibility of characterizing both clump scales within the \emph{same} galaxies. However, in these unlensed galaxies, the proper mapping of the sub-$500\,\mathrm{pc}$ scales is severely limited due to still-insufficient resolution and depth. A complementary effort can therefore be made from the opposite end -- targeting lensed galaxies, where the sub-$500\,\mathrm{pc}$ scales are already resolved, and working upward to characterize the giant clump scales they compose.

Hence, we have started a series of works (\emph{Star-forming Clumps As Layered Emergent Structures}; SCALES) that detects and models the compact lensed clumps (or simply `clumps') within lensed galaxies at $z>1$, while collecting the remaining flux at kpc-scales
into an extended component (`EC'). The EC is not assumed to be a distinct physical structure. It is simply whatever emission is not accounted for by the robustly measured clumps. The implicit assumption is that the clumps and the EC together make up the giant clumps one typically detects in unlensed samples. The first work (\citealt{2026arXiv261003866K}; SCALES-I, S-I hereafter) presented the novel methodological framework, which uses a difference-of-Gaussian image decomposition on the lensed images to simultaneously model the clumps ($\approx 50-400\,\mathrm{pc}$) and the EC ($\sim0.5$--$1\,\mathrm{kpc}$).

In this paper, we ask a simple question: \emph{what is inside a giant clump}? Characterizing the substructures within the giant clumps, and establishing how they relate to their host, will allow us to interpret the marginally resolved sub-$500\,\mathrm{pc}$ structures one might detect in unlensed samples, and to place them within the overall framework of hierarchical star-forming structures. Throughout, we adopt a $\Lambda$CDM cosmology with $\Omega_{\rm m}=0.3$, $\Omega_{\Lambda}=0.7$, and $\mathrm{H_{0}}=70\,\mathrm{km\,s^{-1}\,Mpc^{-1}}$, and a \cite{kroupa01} initial mass function.

\begin{figure}
\centering
\includegraphics[width=0.85\columnwidth]{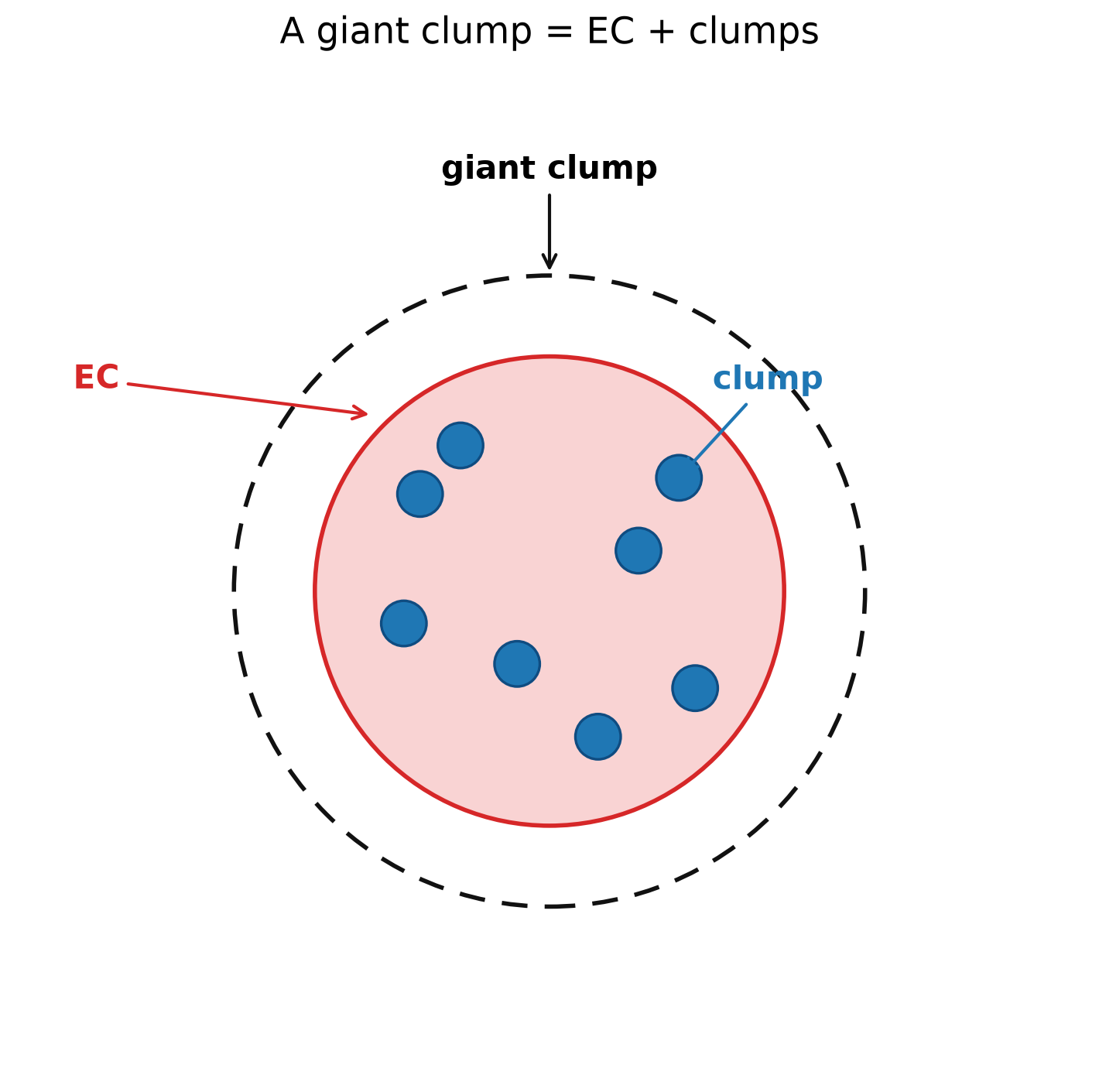}
\caption{Schematic illustration of the terminology adopted in this work. A giant clump (dashed circle) comprises an extended component (EC; red) and the compact clumps (blue) embedded within it.}
\label{fig:giant_clump_schematic}
\end{figure}

Since the distinction between clumps, the EC and giant clumps underlies everything that follows, we reiterate it here at the outset (Fig.~\ref{fig:giant_clump_schematic}). As in S-I, we reserve \emph{clump} for the compact ($\approx 50-400\,\mathrm{pc}$) structures individually resolved in our lensed sample, and \emph{EC} for the surrounding $\sim0.5$--$1\,\mathrm{kpc}$-scale emission left over once these clumps are accounted for, without assuming the EC to be a singular physical structure in its own right. Similar to S-I, the EC therefore contains the light of any undetected clumps. However, unlike in S-I, it also includes the light of the detected clumps associated with it that lack the photometry required for an SED fit of their own (Sec.~\ref{subsec:det_flux_fit}). We note that this definition of \emph{clump} does not encompass the $\sim10\,\mathrm{pc}$-scale star clusters, which we expect to reside at a hierarchical rung below the lensed clumps, as argued in S-I. We reserve the term \emph{giant clump} for the combination of a clump (or clumps) with its associated EC, tied together through the spatial association introduced in Sec.~\ref{subsec:ec_clump_assoc} and illustrated in Fig.~\ref{fig:galaxy_illustration} -- the same object one would recover in an unlensed image at kpc-scale resolution.

\section{Methods} \label{sec:methods}

\subsection{Data and Sample} \label{subsec:data_sample}

Our sample is built upon the most extensive lensed-galaxy census assembled so far \citep[][hereafter C25]{claeyssens25}, centered on the Abell 2744 core. The field is covered by two JWST/NIRCam programs, UNCOVER \citep[GO 2561;][]{bezanson24} and MEGASCIENCE \citep[GO 4111;][]{suess24}, which together provide the broad- and medium-band imaging used here. Following S-I, we retain 13 bands across the short- and long-wavelength channels (F070W, F090W, F115W, F150W, F200W, F277W, F356W, F444W, F140M, F162M, F182M, F210M, and F250M). The redshifts of 18 of the 23 galaxies are spectroscopic \citep{richard21}, while for the remaining five we adopt the UNCOVER photometric redshifts \citep[][see also S-I, Sec.~2.1]{weaver24}. The host galaxy stellar masses and SFRs listed in Table~\ref{tab:host_galaxies} are taken from the C25 catalog.

From the 476 galaxies in C25 we keep those with more than 10 detected clumps, and discard five with close neighbors, leaving a final set of 23 galaxies (Table~\ref{tab:host_galaxies}) at $z=1-4$ with magnifications $\mu = 2.0-9.1$. These are $5\%$ of the C25 hosts, yet enclose 430 C25 clumps ($22\%$ of their catalog). Lensing corrections use the public UNCOVER convergence and shear maps \citep{furtak23,price25}, rescaled to each source's redshift to yield clump-by-clump magnifications, with uncertainties from the MCMC realizations of the lens model that are propagated into the SED-based stellar masses; the full selection and lensing treatment are detailed in S-I (Sec.~2.1, Appendix~A).

\begin{table*}[!htp]
\caption{Host galaxies in the final sample. Stellar masses and SFRs are in log$_{10}$ units ($\mathrm{M_\odot}$ and $\mathrm{M_\odot}\,\mathrm{yr}^{-1}$); $\mu$ is the median of the per-clump lensing magnifications of each galaxy. $N_\mathrm{clump}$ is the number of clumps with a measured stellar mass, $N_\mathrm{EC}$ the number of ECs, and $N_\mathrm{giant}$ the number of giant clumps (an EC plus its associated clumps, requiring at least one associated clump above our mass completeness limit). Quoted uncertainties are the catalog 16th/84th-percentile bounds. Redshifts marked with an asterisk ($^{*}$) are photometric rather than spectroscopic.}
\label{tab:host_galaxies}
\begin{tabular}{l c c c c c c c c c}
\hline\hline
ID & R.A. & Dec. & $z$ & $\mu$ & $\log M_{*}$ & $\log \mathrm{SFR}_{100}$ & $N_\mathrm{clump}$ & $N_\mathrm{EC}$ & $N_\mathrm{giant}$ \\
   & (deg) & (deg) & & & $[\mathrm{M_\odot}]$ & $[\mathrm{M_\odot}\,\mathrm{yr}^{-1}]$ & & & \\
\hline
4304 & 3.59977 & -30.41392 & $1.77$ & $3.44$ & $9.47^{+0.05}_{-0.00}$ & $1.58^{+0.48}_{-0.47}$ & 23 & 5 & 5 \\
7775 & 3.59755 & -30.40388 & $1.69$ & $6.30$ & $9.62^{+0.17}_{-0.04}$ & $0.47^{+0.35}_{-0.08}$ & 22 & 10 & 1 \\
8136 & 3.58285 & -30.40283 & $1.89$ & $8.89$ & $9.09^{+0.11}_{-0.06}$ & $0.21^{+0.25}_{-0.03}$ & 21 & 7 & 3 \\
9198 & 3.57731 & -30.40159 & $2.58$ & $3.81$ & $9.43^{+0.06}_{-0.04}$ & $-0.88^{+0.18}_{-0.14}$ & 7 & 6 & 4 \\
10570 & 3.56888 & -30.39553 & $1.01$ & $1.96$ & $8.41^{+0.03}_{-0.01}$ & $0.44 \pm 0.00$ & 10 & 8 & 1 \\
11188 & 3.57784 & -30.39386 & $1.16$ & $3.71$ & $8.97 \pm 0.08$ & $0.02^{+0.23}_{-0.04}$ & 19 & 7 & 2 \\
11670 & 3.56848 & -30.39230 & $1.54$ & $2.46$ & $9.94^{+0.02}_{-0.01}$ & $1.38^{+0.85}_{-0.88}$ & 28 & 9 & 3 \\
12575 & 3.59146 & -30.38950 & $1.37$ & $3.40$ & $8.95^{+0.08}_{-0.12}$ & $0.86^{+0.47}_{-0.13}$ & 21 & 11 & 3 \\
12800 & 3.56988 & -30.38891 & $3.79$ & $3.79$ & $9.30^{+0.04}_{-0.03}$ & $1.39^{+0.04}_{-0.01}$ & 12 & 3 & 2 \\
13428 & 3.58697 & -30.38700 & $1.86$ & $4.00$ & $8.58^{+0.07}_{-0.13}$ & $0.66^{+0.23}_{-0.04}$ & 8 & 3 & 2 \\
14002 & 3.56501 & -30.38542 & $2.68$ & $3.56$ & $8.46 \pm 0.04$ & $0.51^{+0.01}_{-0.03}$ & 14 & 3 & 2 \\
14601 & 3.55667 & -30.38410 & $2.58$ & $3.58$ & $10.03^{+0.04}_{-0.03}$ & $2.10^{+0.05}_{-0.01}$ & 6 & 2 & 2 \\
15439 & 3.54260 & -30.38268 & $1.41$ & $2.25$ & $10.65^{+0.04}_{-0.03}$ & $0.68 \pm 0.10$ & 76 & 18 & 9 \\
15832 & 3.56231 & -30.38118 & $3.49^{*}$ & $4.58$ & $8.91^{+0.05}_{-0.11}$ & $0.97^{+0.01}_{-0.07}$ & 28 & 4 & 4 \\
16673 & 3.59169 & -30.37959 & $1.80$ & $2.18$ & $8.36^{+0.03}_{-0.04}$ & $0.38^{+0.01}_{-0.00}$ & 13 & 4 & 2 \\
17172 & 3.58441 & -30.37872 & $2.36$ & $2.60$ & $9.62^{+0.04}_{-0.06}$ & $1.69^{+0.08}_{-0.05}$ & 21 & 6 & 5 \\
18098 & 3.54063 & -30.37719 & $1.88$ & $4.03$ & $10.10^{+0.05}_{-0.04}$ & $1.77^{+1.87}_{-1.78}$ & 13 & 5 & 3 \\
19355 & 3.58157 & -30.37488 & $1.85^{*}$ & $2.42$ & $9.28 \pm 0.06$ & $1.38^{+1.11}_{-1.03}$ & 10 & 2 & 2 \\
19708 & 3.55820 & -30.37451 & $2.41$ & $7.40$ & $9.90^{+0.04}_{-0.12}$ & $1.99^{+0.01}_{-0.02}$ & 10 & 9 & 6 \\
20054 & 3.54041 & -30.37505 & $2.58$ & $9.05$ & $10.06 \pm 0.05$ & $0.04^{+0.18}_{-0.19}$ & 23 & 10 & 1 \\
22864 & 3.54791 & -30.36882 & $1.52^{*}$ & $4.38$ & $10.18^{+0.06}_{-0.03}$ & $1.05^{+0.10}_{-0.13}$ & 40 & 13 & 5 \\
23491 & 3.54102 & -30.36718 & $3.49^{*}$ & $6.15$ & $9.79^{+0.07}_{-0.00}$ & $1.88 \pm 0.00$ & 6 & 3 & 2 \\
25706 & 3.53720 & -30.36574 & $1.36^{*}$ & $3.17$ & $10.16^{+0.04}_{-0.01}$ & $2.24^{+0.01}_{-0.00}$ & 20 & 8 & 3 \\
\hline
\end{tabular}
\end{table*}

\subsection{Clumps, ECs and their SED masses} \label{subsec:det_flux_fit}

\begin{figure*}[!htp]
\centering    
\includegraphics[width=0.98\textwidth]{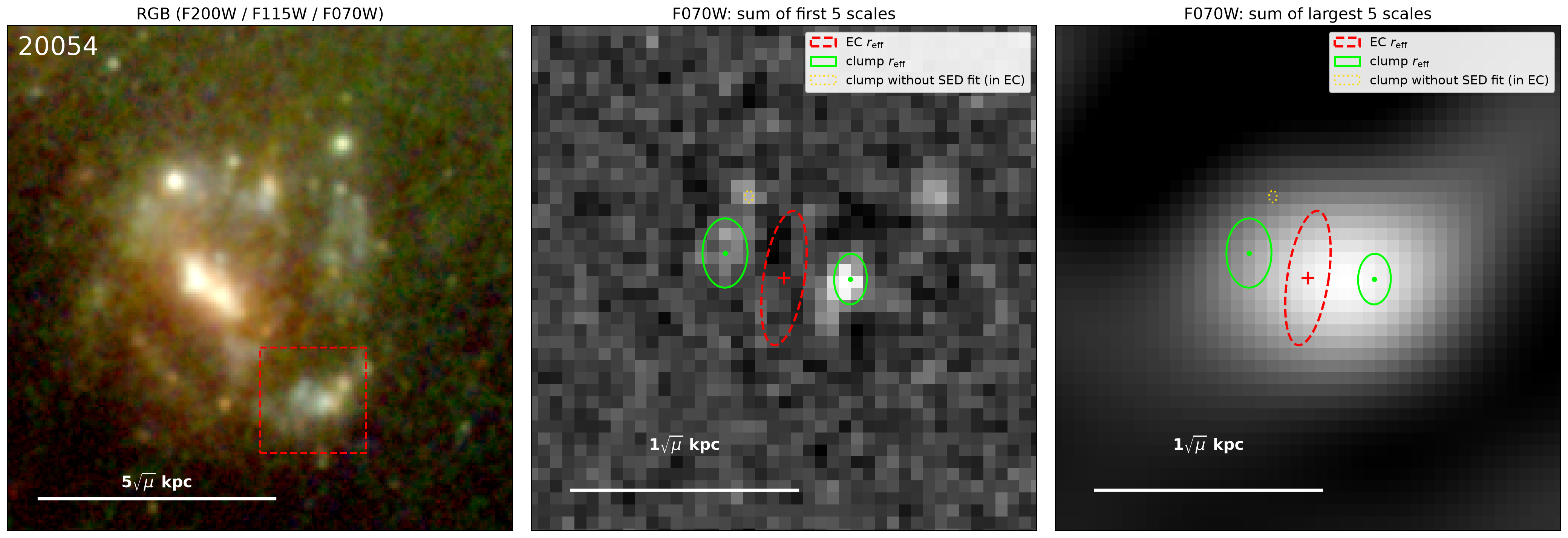}    
\caption{(Left) The RGB (F200W/F115W/F070W) image of galaxy ID: 20054, with one of the giant clumps (EC--clump association) highlighted. (Middle) The same giant clump in an image generated by summing the 5 smallest spatial band-pass planes, with the constituent EC (in red) and the clumps (in green; dotted gold: an associated clump without an SED fit, whose light is fitted with the EC). (Right) The same giant clump in an image generated by summing the 5 largest spatial band-pass planes, with the constituent EC (in red) and the clumps (in green). Each of the EC and clumps is shown as an ellipse (the ellipticity is driven by the lensing shear) where the major axis is the effective radius. The clumps have been associated with the EC since their centers (also shown in the same color) agree within a distance equal to the sum of the effective radii of both components and the F070W PSF FWHM.}
\label{fig:galaxy_illustration}
\end{figure*}

The full methodology is presented in S-I; we summarize only what is needed to follow this work. Clumps are detected and modeled in the F070W reference band (PSF FWHM $=0.029^{\prime\prime}$), chosen for its highest resolution and depth while retaining rest-frame UV coverage across our redshift range. Detection uses a multi-scale difference-of-Gaussian (DoG) decomposition, which isolates structure on a stack of band-pass planes and, in two passes, recovers the compact \emph{clumps} ($r_{\rm eff}\approx 50-400\,\mathrm{pc}$) and the extended components (ECs; $\approx 0.5-1\,\mathrm{kpc}$) at giant clump scales (S-I, Sec.~2.2). As an example, Fig.~\ref{fig:galaxy_illustration} shows the clumps and ECs within a specific region of galaxy ID: 20054. Each is modeled as a PSF-convolved S\'ersic profile with $n=0.5$ (Gaussian), fit simultaneously by minimizing a scale-weighted $\chi^2$ summed over the DoG planes, with clump and EC weights tuned to their respective scales. Fluxes in the remaining 12 bands are obtained by refitting amplitudes alone with the structural variables fixed, but with the 2D models convolved with the PSFs of the respective filters. Estimations of the photometric uncertainties and per-filter completeness limits follow the injection--recovery procedures of S-I (Sec.~2.3).

We note, as in S-I, that the EC is not a physical object but a scale-separated \emph{local background}, absorbing all light that is not a compact clump. These are detected through source selection which prioritizes the largest band-pass planes, thereby detecting clumpy (giant clump-like) structures. Hence, they can consist of truly diffuse emission at $\approx 0.5-1\,\mathrm{kpc}$ scales, sub-threshold clumps, and the overlapping wings of detected clumps alike. \textcolor{blue}{As mentioned earlier, unlike in S-I, the EC also absorbs the flux of any associated clump that is not detected in a sufficient number of filters. Given the lack of a clear physical definition, we emphasize that the EC sizes are estimated properties without robust errors.} 

Stellar masses are derived by fitting the 13-band, magnification-corrected photometry with \texttt{BAGPIPES} \citep{carnall18}, using an exponentially declining SFH, free age and metallicity, \cite{calzetti00} dust, and nebular emission (S-I, Sec.~2.4). We require detections (flux $\geq0.5\times$ its uncertainty) in at least 6 filters for clumps and 4 for ECs (S-I, Appendix~C), and quote posterior-median stellar masses with 16th--84th percentile uncertainties. In this work, we fit each EC together with its associated clumps (Sec.~\ref{subsec:ec_clump_assoc}) that lack the six bands required for an SED fit of their own. We add the fluxes of these clumps to those of the EC in every band, with each component corrected by its own magnification, while keeping the fractional flux uncertainty of the EC. The EC stellar mass is then measured from this summed photometry. Furthermore, we only use the ECs within the F444W host segment. The stellar mass is the only SED-derived quantity we use for the ECs, and throughout our $90\%$ mass completeness limit is $\log(M_\star/\mathrm{M_\odot}) = 7.10^{+0.06}_{-0.06}$ (S-I, Appendix~D).

\subsection{Giant clumps: associating ECs to clumps}
\label{subsec:ec_clump_assoc}

The one aspect specific to this work is the association of clumps with ECs into giant clumps.  A clump is taken to belong to an EC if the distance between their centers is smaller than the sum of their effective radii and the F070W PSF FWHM, with the final term accounting for uncertainties in determining the center of the components,
\begin{equation}
    d(\mathrm{EC},\,\mathrm{clump}) \leq r_{\mathrm{eff}}^{\mathrm{EC}}
    + r_{\mathrm{eff}}^{\mathrm{clump}} + \mathrm{PSF_{FWHM; F070W}}.
\end{equation}
The centers and $r_{\mathrm{eff}}$ are taken from the reference (F070W) fits, and the PSF FWHM ($0.029^{\prime\prime}$) accounts for the resolution limit on separating two centroids.  An EC together with all the clumps associated with it in this manner constitutes a single giant clump. An example of such an association is shown in Fig.~\ref{fig:galaxy_illustration} (further examples can be found in Appendix Fig.~\ref{fig:ec_assoc_collage}). The association is defined per EC, so that a clump lying within reach of more than one EC is assigned to each; this mirrors the situation in unlensed samples, where a compact structure need not belong to a unique giant clump. The giant clump is thus the object one would detect in an unlensed image at kpc-scale resolution: the EC and the compact clumps embedded within it, taken as a whole. Its total stellar mass is the sum of the EC mass and the masses of all its associated clumps.

We acknowledge the aforementioned clump mass completeness limit of our study, and, using it directly ($10^{7.1}\,\mathrm{M_\odot}$), we only accept giant clumps with at least one such clump associated with them in an attempt to propagate the incompleteness to the current study. By this criterion, only 72 of the 156 ECs in our sample ($46\%$) qualify as giant clumps (Table~\ref{tab:host_galaxies}); the remainder are excluded from the giant clump sample entirely. Hence, by definition, the stellar masses of our giant clumps have a lower limit of $10^{7.1}\,\mathrm{M_\odot}$. Including the rest of the ECs in our study would have only resulted in speculative arguments regarding clumps that are below our completeness limit and their associated EC --- components which we effectively treat as part of the same mass budget later in this work.

\section{Results}
\label{sec:results}

The following section details the results based on the stellar mass measurements of the clumps, ECs and giant clumps.

\subsection{Stellar mass budget across components}
\label{subsec:mass_budget}

\begin{figure}     
\centering    
\includegraphics[width=0.46\textwidth]{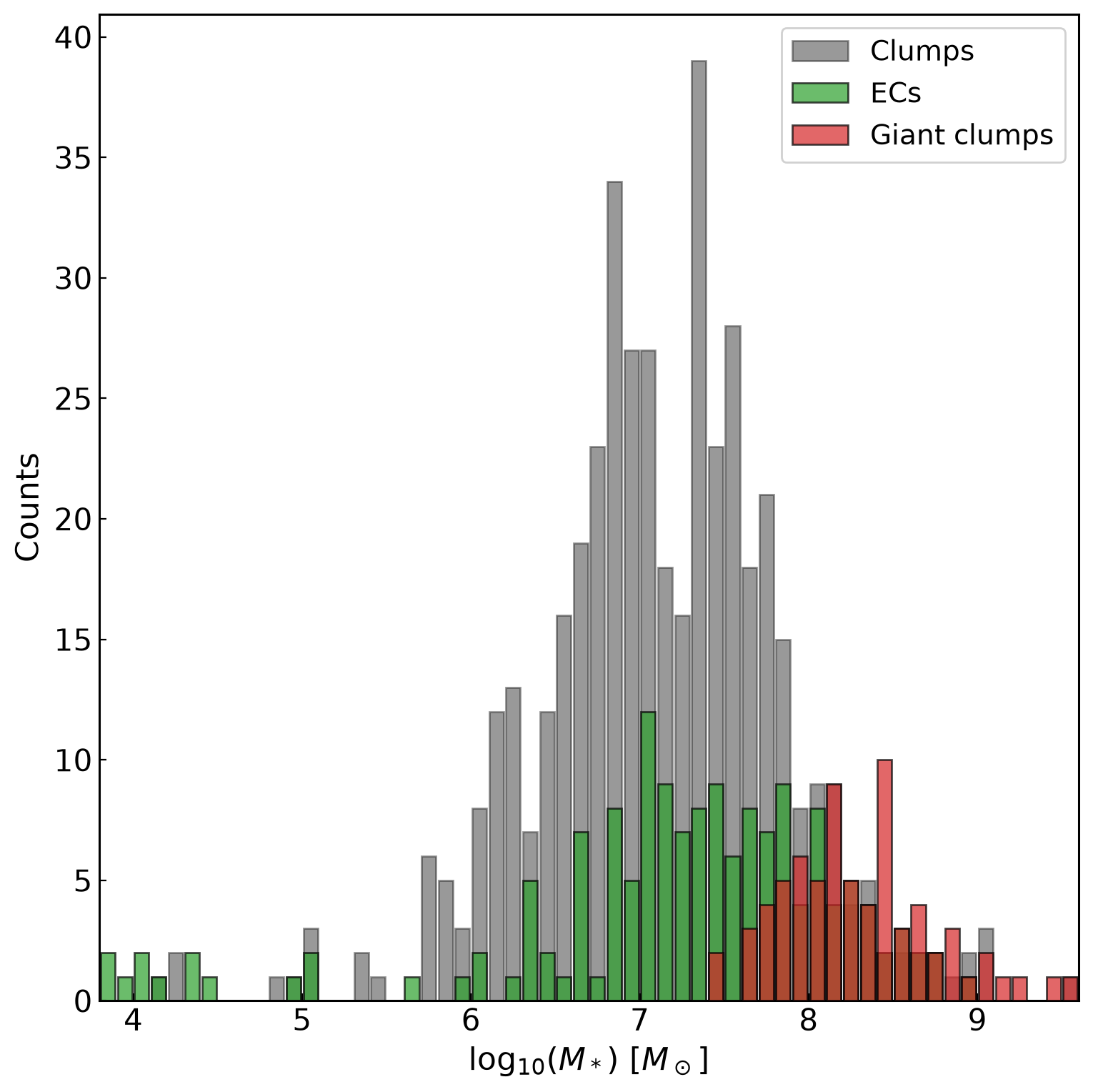}    
\caption{Histogram showing the distribution of stellar mass for clumps, ECs and the giant clumps.}    
\label{fig:mass_histogram}
\end{figure}

\begin{figure}     
\centering    
\includegraphics[width=0.46\textwidth]{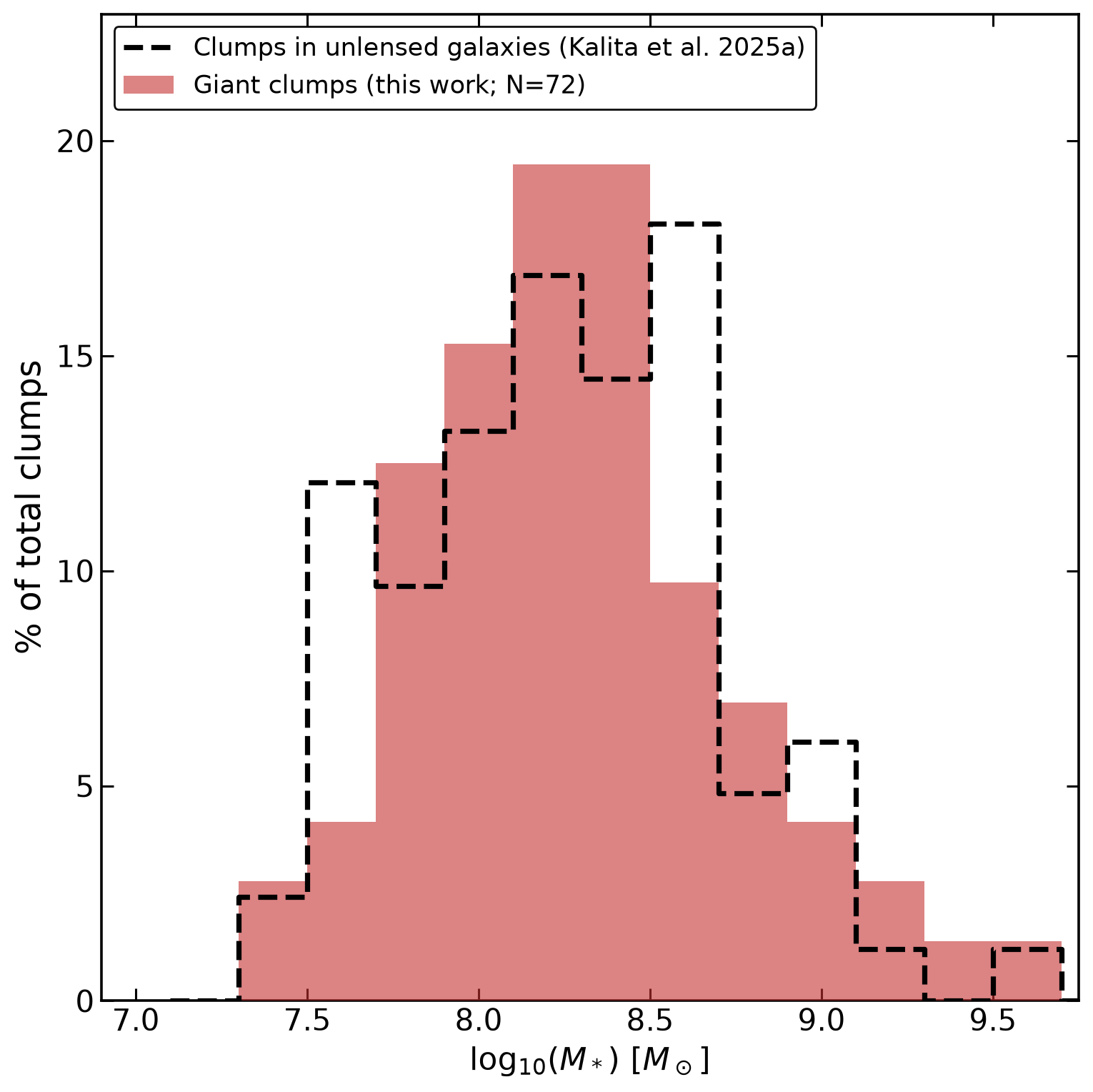}    
\caption{The comparison of the giant clump stellar mass in this work with the clump stellar mass measured in \cite{kalita25a} (unlensed study of clumpy galaxies in the CEERS sample). We only use their clumps in host galaxies with $\log(M_*/\mathrm{M_\odot}) \approx 9.5-10.6$}
\label{fig:mass_comparison}
\end{figure}

As noted in S-I, our sample of 23 galaxies contains 451 clumps and 156 ECs, with both spanning similar stellar mass ranges of $\approx 10^{6}$--$10^{9}\,\mathrm{M_\odot}$ (Fig.~\ref{fig:mass_histogram}). Applying our giant clump selection criteria (Sec.~\ref{subsec:ec_clump_assoc}), we find a total of 72 giant clumps, each with at least one associated clump of stellar mass $\geq 10^{7.1}\,\mathrm{M_\odot}$. For these giant clumps, their total stellar mass is made up of the contributions from the $\geq 10^{7.1}\,\mathrm{M_\odot}$ clumps, $< 10^{7.1}\,\mathrm{M_\odot}$ clumps, and the EC. As in S-I, these giant clumps include the clumps visually identified as possible bulge candidates. We discuss the impact of removing them in Sec.~\ref{subsec:bulge_impact}.

To check the validity of our giant clump definition, which is aimed at mimicking the (giant) clumps reported in unlensed studies, we compare our giant clumps to those measured by \cite{kalita25a}. We only use their clumps in host galaxies with $\log(M_*/\mathrm{M_\odot}) \approx 9.5-10.6$, which matches the mass range of our sample. The two giant clump stellar mass distributions, shown in Fig.~\ref{fig:mass_comparison}, are qualitatively similar. It is worth noting that, unlike their sample, ours is not a stellar-mass-complete sample of galaxies.

\begin{figure}     
\centering    
\includegraphics[width=0.46\textwidth]{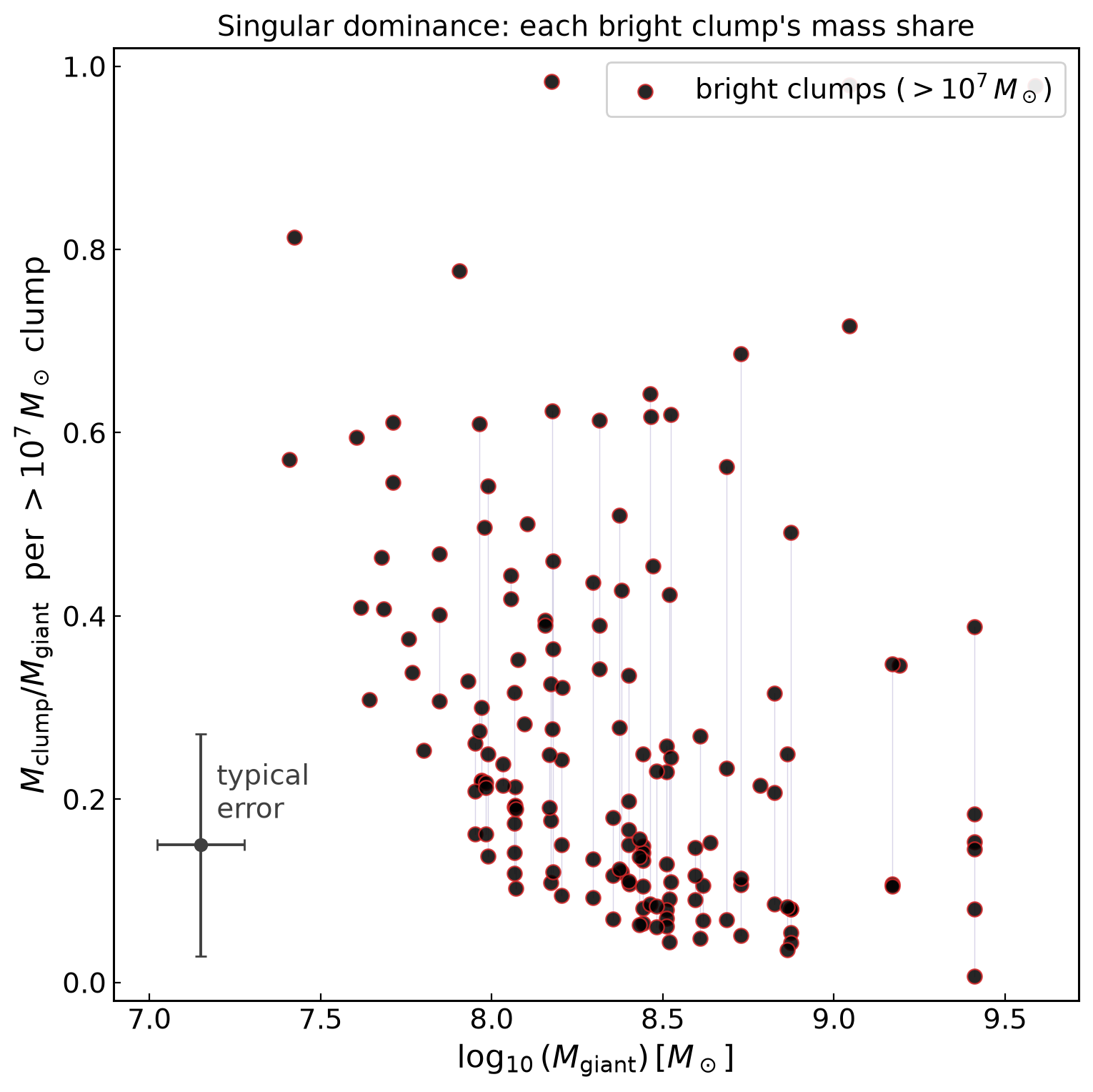}
\caption{The fraction of giant clump stellar mass held by each associated clump with stellar mass $\geq 10^{7.1}\,\mathrm{M_\odot}$, plotted against the giant clump stellar mass. Clumps belonging to the same giant clump are connected by vertical dotted lines.} 
\label{fig:clump_dominant_budget}
\end{figure}

\begin{figure}     
\centering    
\includegraphics[width=0.46\textwidth]{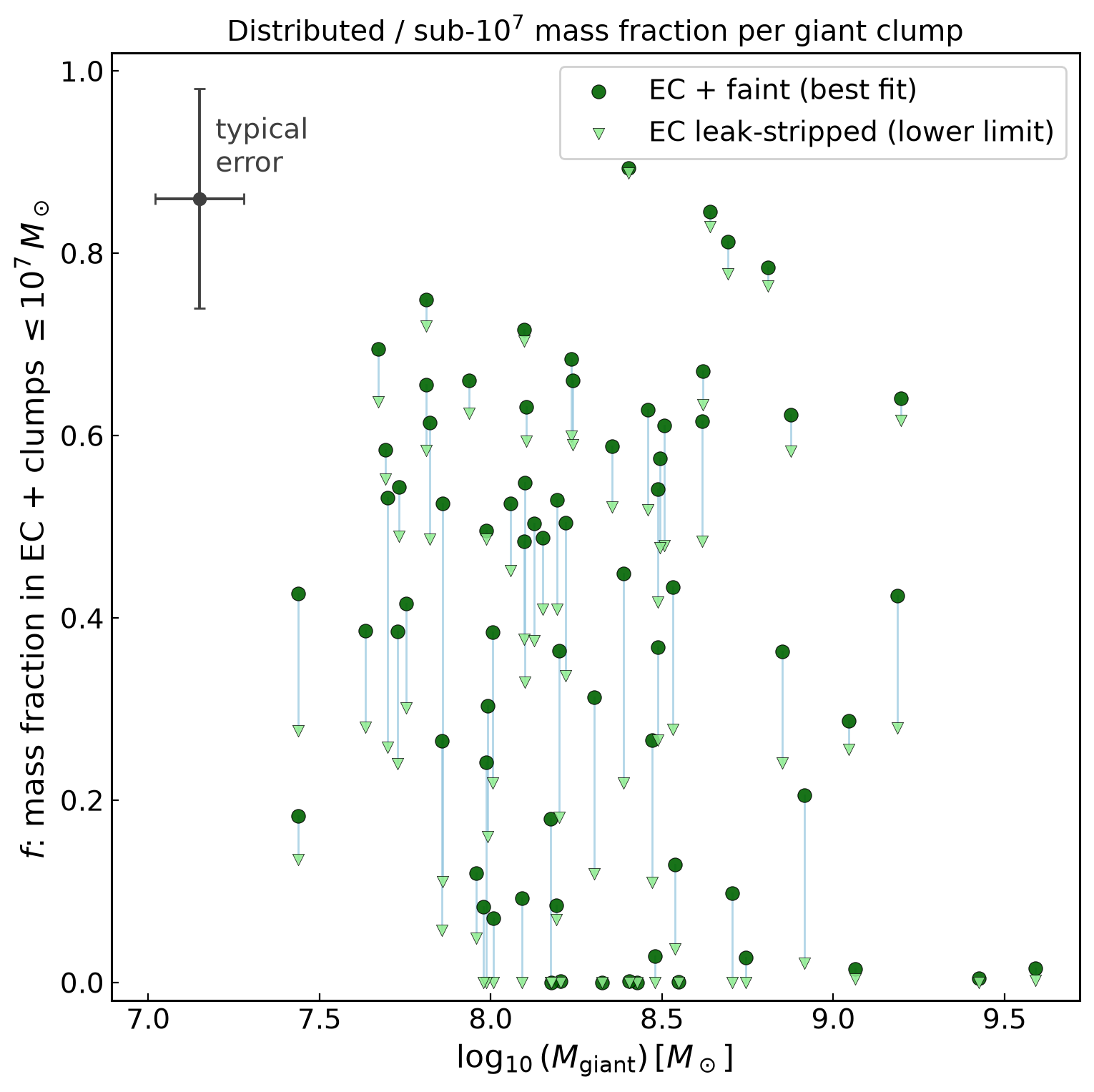}
\caption{The circular markers show the fraction of giant clump stellar mass held by the combination of all clumps $<10^{7.1}\,\mathrm{M_\odot}$ and the EC, plotted as a function of the giant clump stellar mass. The downward-pointing triangles are the lower limits on the mass fraction, accounting for possible contamination of the EC mass by clump wings.}
\label{fig:ec_budget}
\end{figure}

One of our key aims in this work is to guide the interpretation of giant clumps in unlensed galaxy observations that might marginally detect a small fraction of the constituent clumps within, due to a combined effect of resolution and image depth.  As noted in the introduction, \cite{kalita25b} for instance found possible substructure within the giant clumps in their sample.  We argue that whether or not one of the constituent clumps is detected will heavily rely on the fraction of the total giant clump mass budget each of them makes up. Therefore, in Fig.~\ref{fig:clump_dominant_budget}, we show the ratio of clump mass to the host giant clump mass for each of the $\geq 10^{7.1}\,\mathrm{M_\odot}$ clumps it contains. 

A single clump holds more than half of the giant clump mass in 21 of the 72 giant clumps (29\%), and in just 4 cases is this dominance significant at $>1\sigma$.  Here the uncertainty ($\sigma$) is based on the errors in the stellar mass of each of the components being propagated into this assessment. Among the 41 giant clumps hosting at least two clumps above $10^{7.1}\,\mathrm{M_\odot}$, the two most massive clumps together hold more than half of the giant clump mass in 23 cases (56\%), and in 8 cases (20\%) if one requires this to be significant at $>1\sigma$.

We finally explore the instances in which the extended component (EC), together with the faint ($<10^{7.1}\,\mathrm{M_\odot}$) clumps, accounts for more than half of the giant clump mass (Fig.~\ref{fig:ec_budget}).  We deliberately group the EC and the $<10^{7.1}\,\mathrm{M_\odot}$ clumps together, since we are dealing with an incomplete sample of clumps in this regime, while the EC is expected to be a flux sink of anything not associated with a clump.  Therefore, separating these components places over-reliance on a decomposition that is not expected to be robust.  We further explore this in Sec.~\ref{subsec:mass_function}.

We find this combined component of EC and $<10^{7.1}\,\mathrm{M_\odot}$ clumps to be dominant in 31 of the 72 giant clumps ($43\%$). One concern, however, is that some of the EC flux entering this calculation may be the extended wings of the clumps that have not been properly modeled. We therefore carry out an experiment to place a highly conservative lower limit on the fraction shown in Fig.~\ref{fig:ec_budget}. We inject all clumps with their measured parameters into an empty sky patch, creating a clump-only image of each galaxy, and then repeat the full measurement analysis while holding the positions of all clumps and ECs fixed to those from the original run. By construction, this forces the EC in every galaxy to measure flux that is not truly present. We stress that this does not estimate the leakage itself, but rather sets an aggressive upper bound on its possible value. The stellar mass recovered in this way is \emph{assumed} to originate entirely from the wings of the $\geq 10^{7.1}\,\mathrm{M_\odot}$ clumps, making the estimate even more conservative, and we subtract it from the combined EC and faint-clump mass entering the ratio that defines their contribution to the giant clump. This yields the highly conservative lower limit shown by the triangles in Fig.~\ref{fig:ec_budget}, for which the number of giant clumps dominated by this component drops to 18 ($25\%$).

\subsection{The CSMF and modifications}
\label{subsec:mass_function}

\begin{figure}     
\centering    
\includegraphics[width=0.465\textwidth]{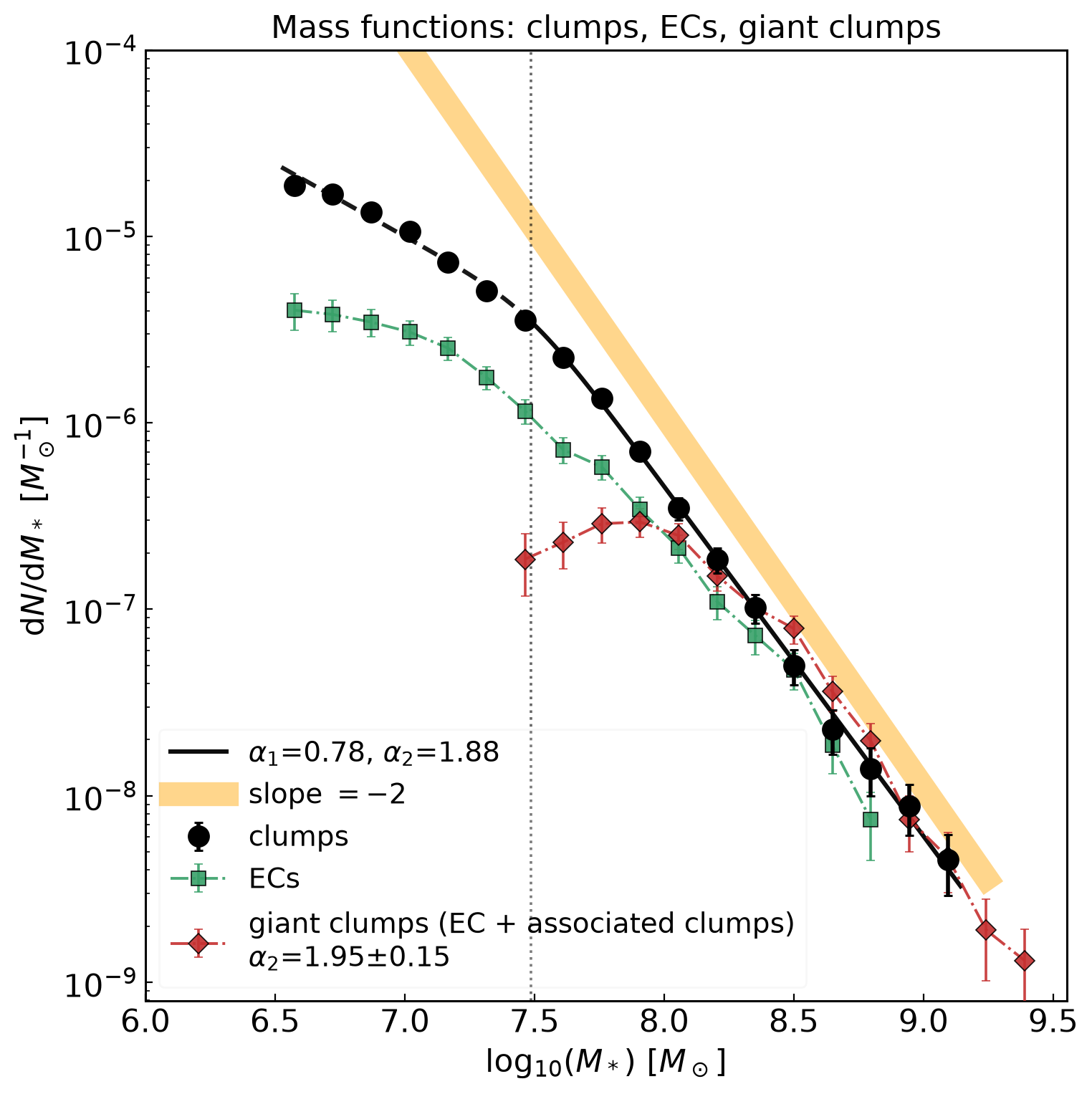}   
\includegraphics[width=0.45\textwidth]{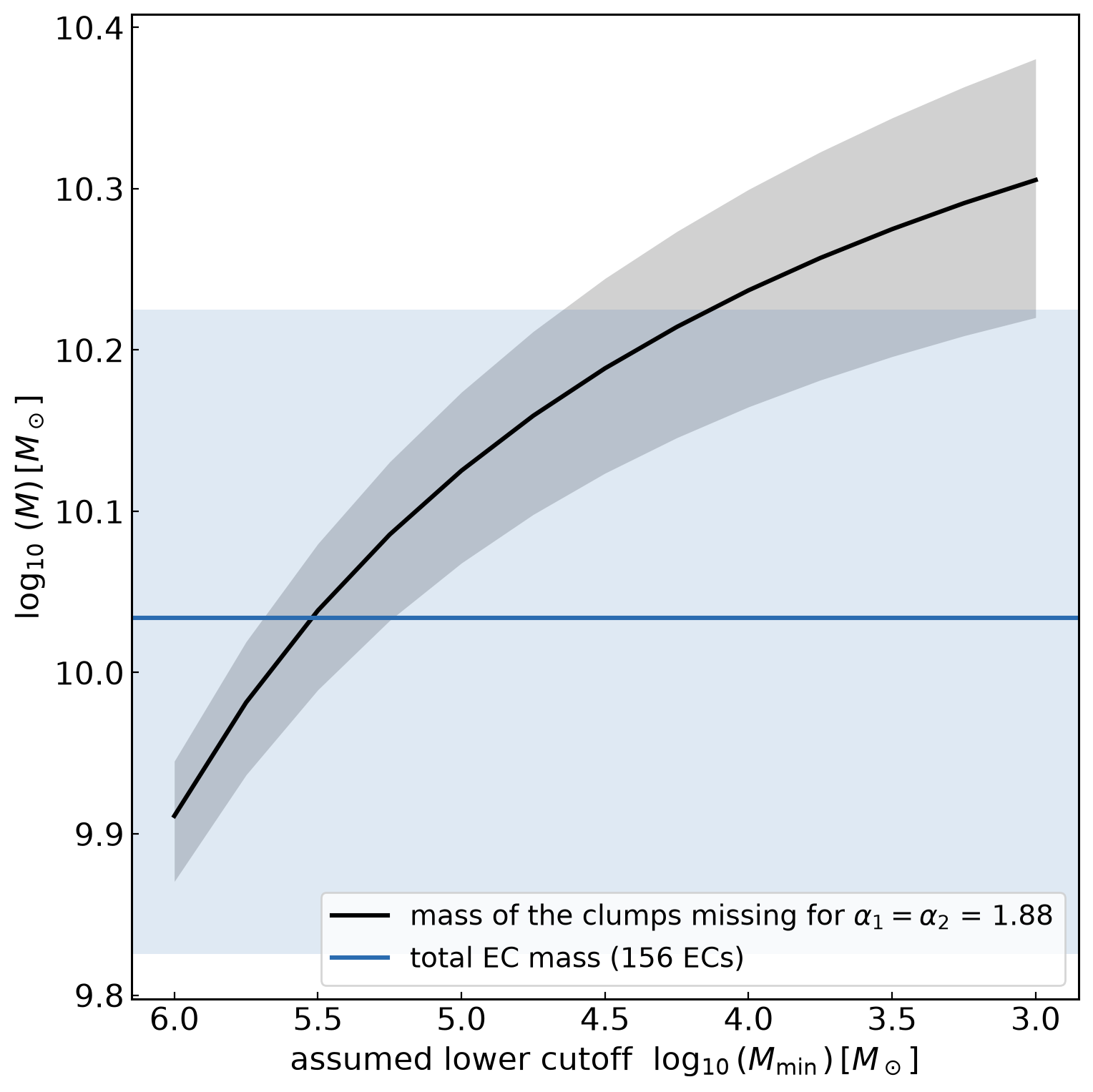} 
\caption{(Top) The clump, EC and giant clump stellar mass functions, showing the number of objects per unit stellar mass ($dN/dM$) as a function of stellar mass. The vertical dotted line shows the best-fit break value for the CSMF; the giant clump mass function is fit with its own break (not shown), while the ECs are not fit. (Bottom) The stellar mass that the undetected clumps would need to contain for the CSMF to continue below the break with the high-mass slope ($\alpha_1=\alpha_2=1.88$), plotted as a function of the assumed lower cutoff mass (black). This is the mass of the clumps missing between the observed CSMF and its extension, with the shaded region showing the $\pm1\sigma$ uncertainty in $\alpha_2$. The blue line shows the total stellar mass of the 156 ECs, with the shaded region spanning their 16th--84th percentile masses.}
\label{fig:mass_function}
\end{figure}

One key result of S-I was the clump stellar mass function (CSMF), whose slope allows population-level comparisons to star-forming structures, like star clusters, detected in the local Universe \citep[e.g.][]{zhang99, gieles06a, gieles06b, dowell08, larsen09, adamo17}. The CSMF was found to be represented by a broken power law, with a slope of $1.88 \pm 0.04$ above a clump stellar mass of $\approx 10^{7.49}\,\mathrm{M_\odot}$. This is shown in Fig.~\ref{fig:mass_function} (upper panel; with details regarding its construction found in S-I). Below this mass, the slope is $0.78 \pm 0.05$. Although the median clump detection ($90\%$) completeness of our study across the clump sample sits at $\log(M_\star/\mathrm{M_\odot}) = 7.10^{+0.06}_{-0.06}$, which is below the break, this threshold is heavily dependent on clump properties (e.g., stellar age, as shown in Appendix Fig.~21 of S-I). Hence, one cannot directly interpret the effect of completeness on the CSMF.

However, one key feature of our study is that we include any flux not associated with the detected clumps in the EC. Hence, a pertinent question is whether the break disappears if this additional mass is included in the CSMF. We test this using the stellar mass budget, where we treat the EC mass as a reservoir of undetected clumps. For this purpose, we extend the CSMF above the break, $dN/dM\propto(M/M_b)^{-\alpha_2}$, down to a lower cutoff mass ($M_{\rm min}$), i.e., we set $\alpha_1=\alpha_2$. We then compare the number of clumps predicted by this extension with the observed number, in $0.15$\,dex bins between $M_{\rm min}$ and $M_b$. As in S-I, we treat each clump as a distribution of 100 points drawn from its stellar mass uncertainty. The difference in each bin, multiplied by the mass at the bin center, gives the mass of the missing clumps. We compare this to the total stellar mass of the 156 ECs, which also includes the clumps without an SED fit of their own (Sec.~\ref{subsec:det_flux_fit}). The uncertainty of the missing mass is given by the 16th--84th percentile range over 300 draws of $\alpha_2$ within its uncertainty, while that of the EC mass is given by the sum of the 16th and 84th percentile masses of the ECs. Since the mass in a power law with a slope below 2 increases as $M_{\rm min}$ decreases, we repeat this comparison for $M_{\rm min}=10^{6}$--$10^{3}\,\mathrm{M_\odot}$.

We find the clump population below the break to be missing a mass of $\log(M_\star/\mathrm{M_\odot}) = 9.9 - 10.3$ for cutoff masses of $\log(M_{\rm min}/\mathrm{M_\odot}) = 6.0 - 3.0$. Meanwhile, the ECs provide a total mass of $\log(M_\star/\mathrm{M_\odot}) = 10.0 \pm 0.2$. Hence, we find that the CSMF can retain the measured high-mass slope down to $\approx 10^{5.5}\,\mathrm{M_\odot}$ if we only consider the median values, and down to $\approx 10^3\,\mathrm{M_\odot}$ when accounting for the uncertainties. However, this test does not account for the dependence of our completeness limit on the clump age. In S-I (Appendix~D), the $90\%$ completeness limit of $10^{7.10}\,\mathrm{M_\odot}$ applies to the clumps at their observed ages (median $0.28\,\mathrm{Gyr}$), but increases to $10^{7.46}$ and $10^{8.33}\,\mathrm{M_\odot}$ if the same clumps are aged to $0.3$ and $1.1\,\mathrm{Gyr}$, respectively. Hence, the CSMF and its extension below the break describe the detectable, and mostly young, clumps, while older clumps can be missing even above the break.

Since we include all the flux not associated with a detected clump in the EC, the mass of these older clumps ($>0.3\,\mathrm{Gyr}$), if they lie within the confines of our models, is still included in our measurements, but as part of the EC. It is unclear how their presence would alter the results, and establishing their effect is beyond the scope of the current observations. Furthermore, we cannot reject the possibility of the CSMF having an intrinsic break, with the clumps below it simply losing their structure immediately after formation and thus adding to the EC mass. This would mean our mass budget estimate refers to the initial mass function of the clumps rather than after it has been allowed to evolve. Hence, we emphasize that through this exercise we simply wish to convey that the detection of the break does not allow for a clear interpretation, but the detection completeness is likely a major contributory factor.

Finally, we fit a similar broken power law to the giant clump population, with the break left free. We find a break at $10^{8.20\pm0.08}\,\mathrm{M_\odot}$ and a high-mass slope of $1.95 \pm 0.15$, consistent with the high-mass slope of the CSMF. This is also consistent with the slope of $1.85\pm0.19$ measured for the giant clumps in unlensed galaxies by \cite{kalita25b} \citep[see also][]{kalita25a, zhu26}. As discussed in S-I (Sec.~4.1.2), their fit only covers clumps above $\approx10^{8.7}\,\mathrm{M_\odot}$ in massive ($>10^{10.5}\,\mathrm{M_\odot}$) main-sequence galaxies at $z\approx1.5$. Hence, the clumps above $M_b$, the giant clumps they make up, and the unlensed (giant) clumps all show slopes consistent with each other, as well as with the fiducial slope of $\approx 2$ expected from scale-free, hierarchical star formation (S-I). This gives the same scale-free mass function slope at (lensed) clump and giant clump scales, although the two populations are nested, with the clumps being the constituents of the giant clumps (Sec.~\ref{subsec:size_dominance_test}). However, as noted in S-I, the slope alone cannot tell whether giant clumps are random blends of the smaller clumps or physically associated clump clusters. Meanwhile, the break in the giant clump mass function is likely shaped by our selection, since each giant clump must host at least one clump above $10^{7.1}\,\mathrm{M_\odot}$. It is worth noting that we do not fit a power law to the EC mass function. The EC is a sink for all the emission not assigned to a detected clump (Sec.~\ref{subsec:det_flux_fit}), rather than a clear physical entity, and hence its mass function does not have a direct physical interpretation.

\section{DISCUSSION}
\label{sec:discussion}

In this work, we have assessed the internal structure of the $>500\,\mathrm{pc}$ giant clumps that are routinely detected in unlensed studies. We have divided the constituents into two broad groups: (I) The clumps (with typical sizes $\approx 50-400\,\mathrm{pc}$) with stellar mass higher than the completeness limit of our study ($10^{7.1}\,\mathrm{M_\odot}$). (II) The rest of the clumps ($< 10^{7.1}\,\mathrm{M_\odot}$) along with the EC. Collectively, they make up the giant clumps through an EC and clump association established using spatial proximity (Sec.~\ref{subsec:ec_clump_assoc}).  The two main results have focused on the distribution of the mass budget of the giant clumps across these components (Sec.~\ref{subsec:mass_budget}), and the nature of the CSMF that can describe the combined population (Sec.~\ref{subsec:mass_function}). In the following section, we discuss the physical implications of these results.

\subsection{The internal composition of a giant clump}
\label{subsec:stochastic_composition}

\begin{figure*}[!htp]
\centering   
\includegraphics[width=0.49\textwidth]{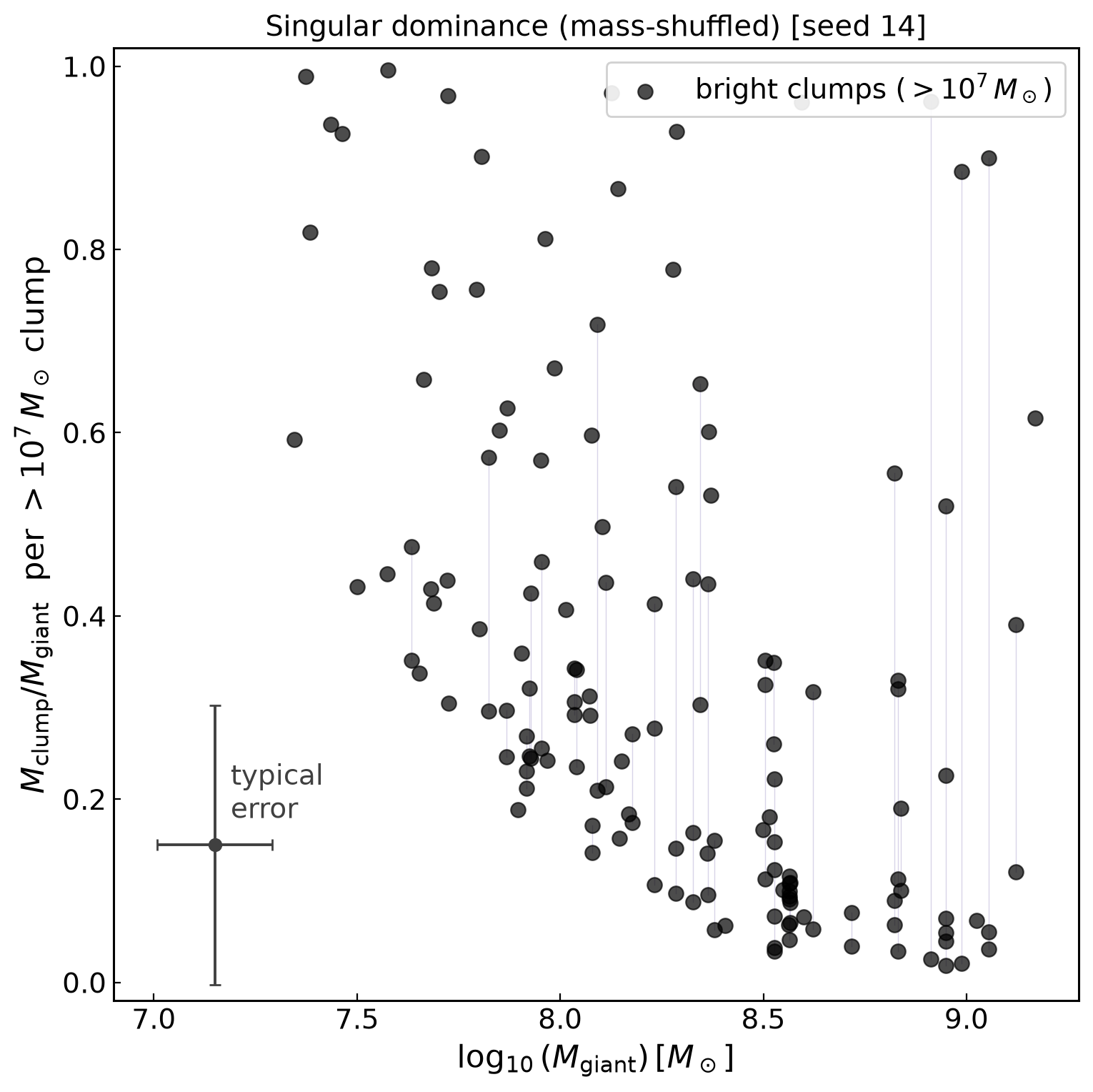} 
\includegraphics[width=0.49\textwidth]{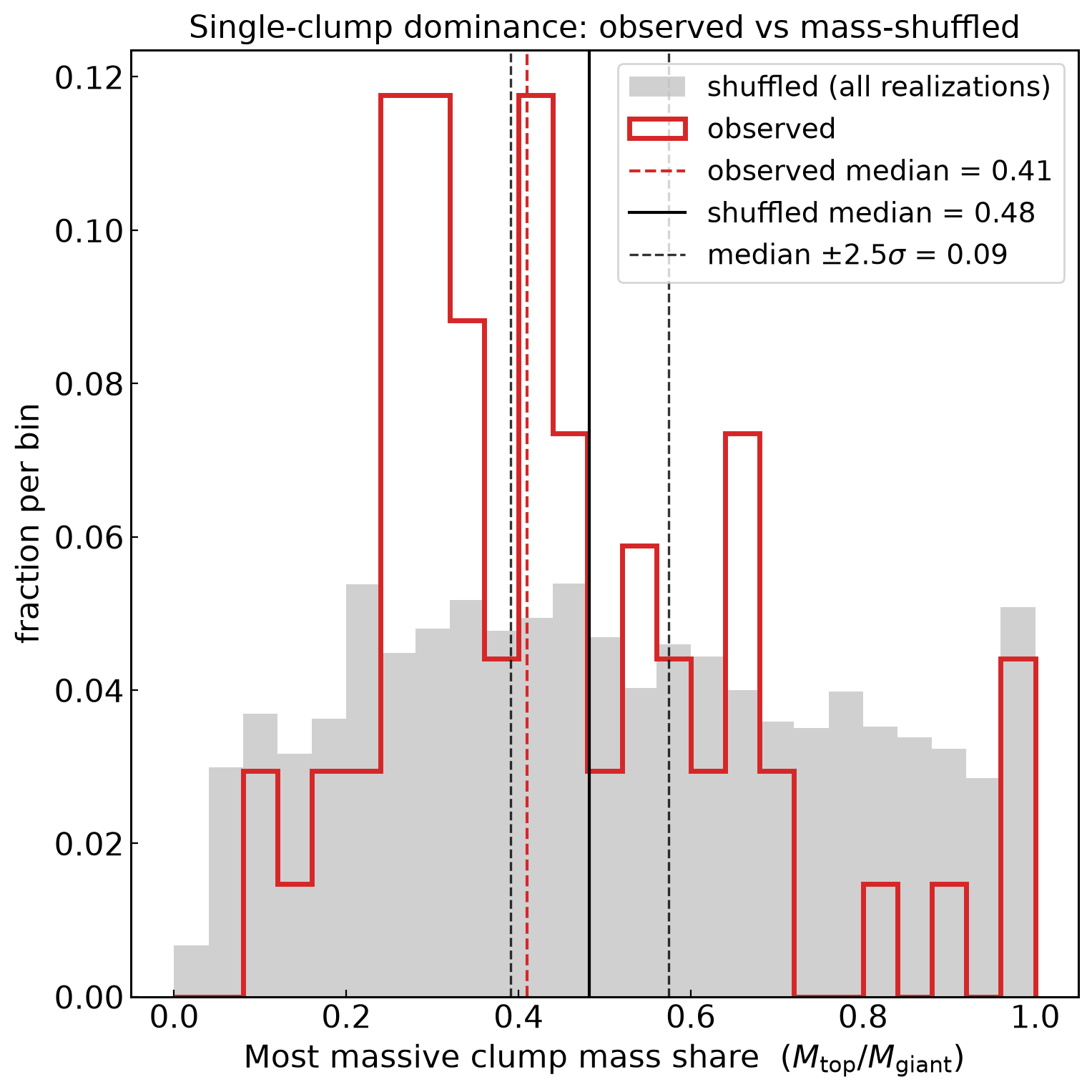}
\caption{(Left) Recreation of Fig.~\ref{fig:clump_dominant_budget}, but with the clump masses in each galaxy reshuffled. The EC--clump associations that make up the giant clumps are kept frozen. (Right) The same process is repeated 500 times, and the share of giant clump stellar mass held by the most massive constituent clump is recorded in each run. The distribution of this fraction is plotted as a histogram. The same fraction for the observed giant clump sample is shown in red. The medians of each distribution are shown as vertical lines of the same color, with the $2.5\sigma$ scatter of the median across the shuffled runs shown as dashed gray lines.}
\label{fig:shuffle_test}
\end{figure*}

\begin{figure}     
\centering    
\includegraphics[width=0.46\textwidth]{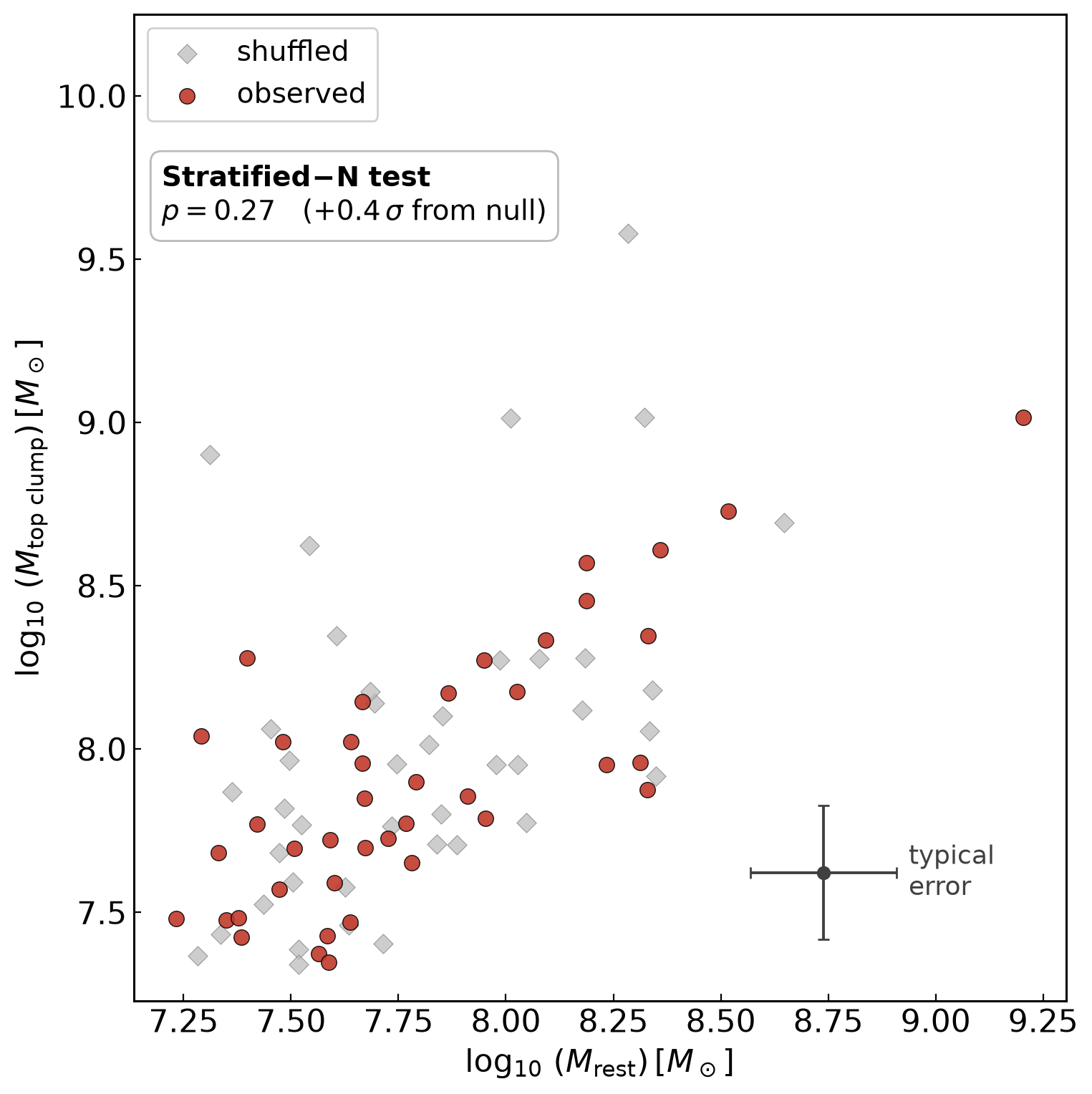} 
\caption{The mass of the most massive clump as a function of the summed mass of the remaining associated clumps, for each giant clump in our sample, shown in red. The gray dots are the same quantities after reshuffling the stellar masses of all clumps within a galaxy, while keeping the EC--clump associations that make up the giant clumps frozen. Only one of the 2000 realizations of this shuffling exercise is shown. The $p$-value quoted at the top left is a metric distinguishing the two populations.}  
\label{fig:max_mass_plot}
\end{figure}

In Sec.~\ref{subsec:mass_budget}, we found that in $29\%$ of giant clumps, more than half of the mass is contributed by the single most massive constituent clump.  Now, we attempt to use the fractional dominance of the most massive clump as a tracer of whether the stellar mass of the giant clump influences the stellar mass of its constituents, thereby indicating a causal link.  Keeping the EC--clump associations fixed, we reshuffle the stellar mass of all clumps within each galaxy. In other words, the stellar mass of each clump is randomly assigned from the full set of stellar mass values of all clumps in the host galaxy. This preserves the CSMF, the number of clumps, and the association geometry, while breaking any link between where a clump lies and how massive it is.  We then regenerate Fig.~\ref{fig:clump_dominant_budget} with this clump-mass-shuffled run and we provide it in Fig.~\ref{fig:shuffle_test} (left panel). Visually, both plots show remarkable similarity.

Repeating this process 500 times, we measure the distribution of the stellar mass share held by the most massive clump ($M_{\rm top}$) in each giant clump during each run. We show the combined distribution in gray in Fig.~\ref{fig:shuffle_test} (right panel), measuring the median value in each run. The combined median and $2.5\sigma$ scatter of the medians of the 500 runs are also estimated. We then compare this to the same mass share in our observed sample. We find the observed and shuffled medians, $0.41$ and $0.48$, to differ by 0.07, which lies within the $2.5\sigma$ scatter of the shuffled medians ($2.5\sigma = 0.09$). Hence, it provides no evidence for giant clumps preferentially hosting a dominant clump, beyond what random sampling of the CSMF already produces. 

The mass fraction, however, reduces each giant clump to a single number. We therefore also check the mass of $M_{\rm top}$ against the summed mass of its companion clumps ($M_{\rm rest}$) (Fig.~\ref{fig:max_mass_plot}). We use the companion mass here rather than the total giant clump mass, since the latter contains $M_{\rm top}$ itself and would therefore correlate the two axes by construction. We also exclude the EC for fear of any leakage from $M_{\rm top}$ driving any correlation, ensuring $M_{\rm top}$ and $M_{\rm rest}$ share no mass in common. We only use the clumps above our completeness limit ($M_\star\geq10^{7.1}\,\mathrm{M_\odot}$) for $M_{\rm top}$, $M_{\rm rest}$ and the shuffles. Hence, this test uses the 41 giant clumps hosting at least two such clumps.  Once again, we redo this measurement by mass-shuffling the clumps 2000 times (with one realization shown in Fig.~\ref{fig:max_mass_plot} for reference).

To compare the observed and shuffled distributions of these two quantities, we use the energy distance \citep{szekely13}, a measure of the separation between two sets of points that requires no binning and makes no assumption about their shape.
For two sets of points $O$ (observed) and $S$ (shuffled), it is defined as
\begin{equation}
\mathcal{E}(O,S) = 2\,\overline{\| o - s \|} - \overline{\| o - o' \|} - \overline{\| s - s' \|},
\end{equation}
where $\overline{\| o - s \|}$ is the mean distance between points of $O$ and $S$. The remaining terms are the mean distances between pairs of points within $O$ and within $S$, where $o'$ and $s'$ denote another point of each set. $\mathcal{E}$ is zero when the two sets are drawn from the same distribution and positive when they differ.

Applied to $M_{\rm top}$ and $M_{\rm rest}$ directly, this comparison would be sensitive to the number of associated clumps $N$: a giant clump with more clumps has both a larger value for its most-massive clump, because the largest of many draws is larger, and a larger companion mass, because more clumps are summed, so $M_{\rm top}$ and $M_{\rm rest}$ rise together even when the masses are assigned at random. To remove this effect, we evaluate $\mathcal{E}$ in a stratified form. We divide the giant clumps into groups of fixed $N$ across the sample,  and merge adjacent $N$ groups so that each group contains at least six giant clumps, which divides the full sample into three groups. We then compute $\mathcal{E}$ separately within each group, and combine the groups into a single statistic weighted by the number of giant clumps in each,
\begin{equation}
\mathcal{E}^{\mathrm{strat}} = \sum_k \frac{n_k}{N_{\rm tot}}\, \mathcal{E}(O_k, S_k),
\end{equation}
where $O_k$ and $S_k$ are the observed and shuffled points in the $k$-th group, $n_k$ is the number of giant clumps it contains, and $N_{\rm tot}$ is the total. Since every term compares giant clumps at fixed $N$, the number of clumps cannot drive the statistic.

However, since the observed and shuffled samples are built from the same clumps and are therefore not independent, one cannot use $\mathcal{E}^{\mathrm{strat}}$ to draw any conclusions. We overcome this by undertaking a two-step procedure. We take one of the shuffled runs as a reference and compare the observed set to it to obtain a value $\mathcal{E}^{\mathrm{strat}}_{\mathrm{obs}}$. We then replace the observed set with each of the remaining shuffled runs to obtain a series of values $\mathcal{E}^{\mathrm{strat}}_{\mathrm{shuffle},i}$ (where $i$ runs from $1$ to $1999$). We define the $p$-value as the fraction of $\mathcal{E}^{\mathrm{strat}}_{\mathrm{shuffle},i}$ values that exceed $\mathcal{E}^{\mathrm{strat}}_{\mathrm{obs}}$. Measurement uncertainties are folded in by drawing each mass from its posterior within every realization. We find the observed points to be indistinguishable from the shuffled ones with $p = 0.27$.  A value $<0.05$ would have indicated the observed configuration to be distinct from the shuffled runs. $\mathcal{E}^{\mathrm{strat}}_{\mathrm{obs}}$ is found to be only $0.4\sigma$ above the mean of all $\mathcal{E}^{\mathrm{strat}}_{\mathrm{shuffle},i}$. Hence, we conclude that the observed configuration of the clumps above $10^{7.1}\,\mathrm{M_\odot}$ is consistent with any random reassignment of their stellar masses. 

\begin{figure*}
\centering
\includegraphics[width=\textwidth]{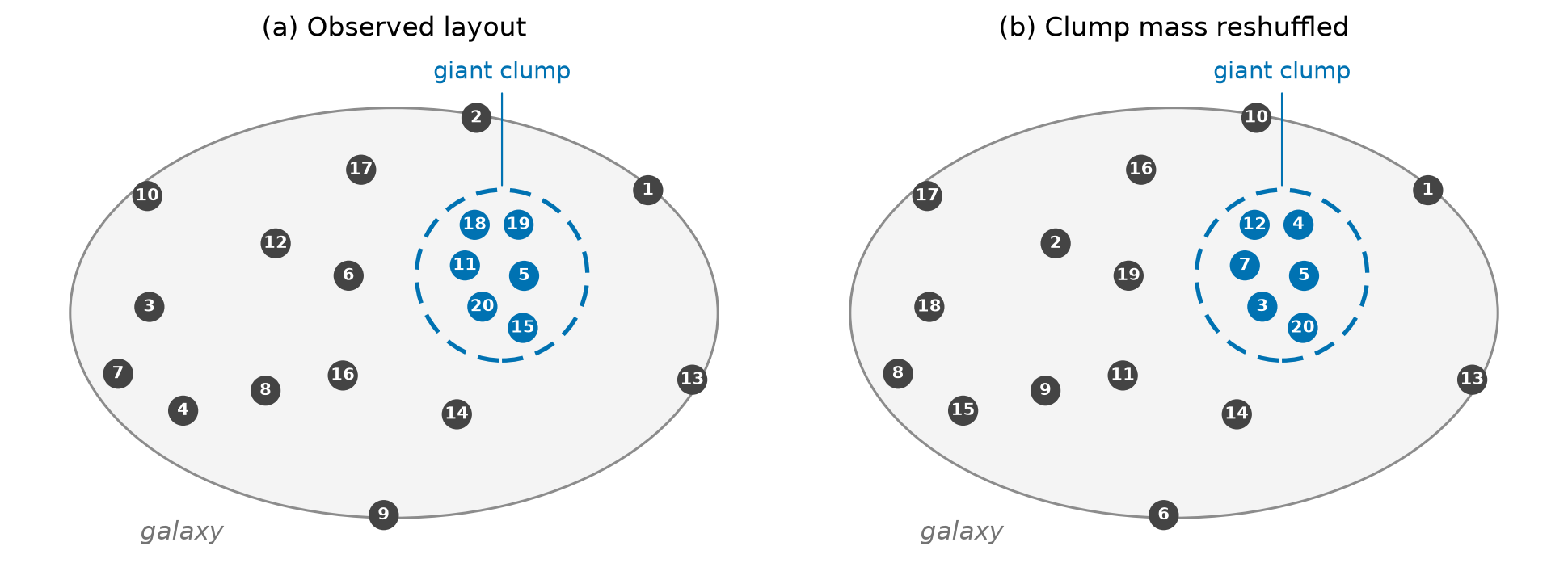}
\caption{Schematic illustration of the composition of giant clumps. (a) A galaxy (gray ellipse) containing 20 clumps with different stellar masses, assigned numbers for identification. A clustered group of clumps is enclosed by a dashed circle, representing a giant clump. (b) The same galaxy and clump positions, but with the clump labels (and hence their stellar mass) reshuffled. Our results suggest that the mass composition of the giant clump in both cases is effectively identical, at least for the clumps above our completeness limit.}
\label{fig:composition_schematic}
\end{figure*}

Using both these tests, we arrive at the same conclusion: within our measurement uncertainties, the stellar mass distribution of the clumps above our completeness limit ($10^{7.1}\,\mathrm{M_\odot}$) inside a giant clump can be explained through random draws from the clump population within its host galaxy. Hence, more massive giant clumps are simply made up of more clumps, and \emph{their composition, at least above $10^{7.1}\,\mathrm{M_\odot}$, can be explained through random sampling of the CSMF}. It is worth noting that we cannot establish whether this also extends to the clumps below our completeness limit. Within these limits, the mass of the most massive clump carries no information about the giant clump it belongs to, beyond simple statistics that would dictate how massive the clump can be. This is illustrated schematically in Fig.~\ref{fig:composition_schematic}. However,  it is worth noting that our assessment is limited by the uncertainties of our measurements. Any additional internal effects might simply be washed out by the large uncertainties. With only 41 giant clumps entering the stratified test, split into three $N$-stratified groups of 15, 20 and 6, our power to detect a weak physical ordering of the clump masses could be improved with a larger sample.

\subsection{The SCALES model: The giant clump composition and the bridge between lensed and unlensed observations}
\label{subsec:size_dominance_test}

\begin{figure*}
\centering
\includegraphics[width=\textwidth]{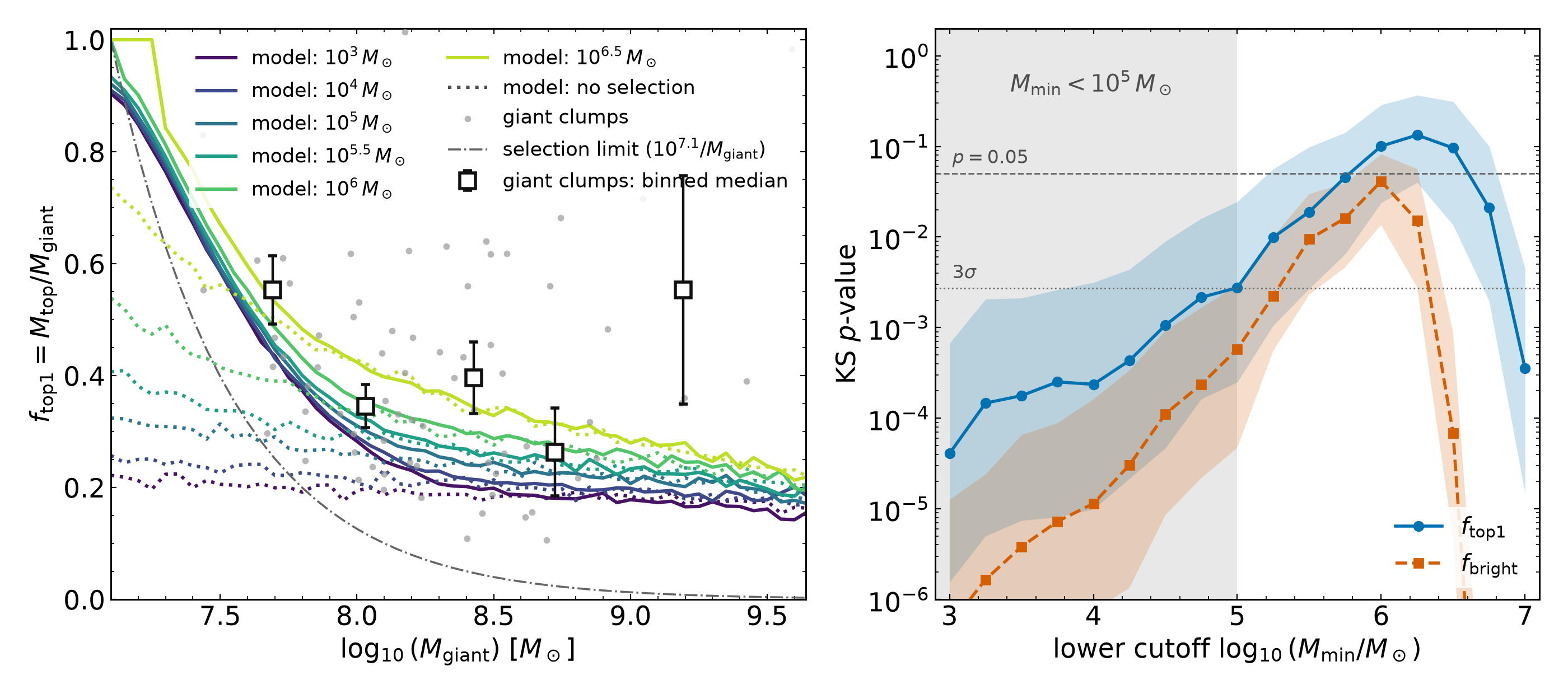}
\caption{(Left) The fractional stellar mass held by the most massive constituent clump, $f_{\rm top1}\equiv M_{\rm top}/M_{\rm giant}$, as a function of giant clump stellar mass. The gray points show the 72 giant clumps in our sample, and the open squares show their median in bins of giant clump mass (with bootstrapped $1\sigma$ uncertainties). The colored lines show the median predicted by the SCALES model for lower cutoffs of the CSMF between $10^{3}$ and $10^{6.5}\,\mathrm{M_\odot}$, with the same selection as for our sample (Sec.~\ref{subsec:size_dominance_test}), while the dotted lines of the same color show the corresponding medians without this selection. The dash-dotted line marks the lower limit on $f_{\rm top1}$ imposed by this selection. (Right) The median $p$-value of the Kolmogorov--Smirnov (KS) test comparing the observed giant clumps with the distribution predicted by the SCALES model at their respective masses as a function of the lower cutoff of the CSMF. The colored bands show the 16th--84th percentile range of the $p$-value. This is shown for $f_{\rm top1}$ (blue) and for the fraction of giant clump stellar mass held by the clumps above $10^{7.1}\,\mathrm{M_\odot}$ ($f_{\rm bright}$; orange). The dashed and dotted horizontal lines mark $p=0.05$ and $p=0.0027$ ($3\sigma$), respectively, and the gray shaded region marks the lower cutoffs below $10^{5}\,\mathrm{M_\odot}$.}
\label{fig:dominance_and_ratio}
\end{figure*}

In Sec.~\ref{subsec:stochastic_composition} we demonstrated that the stellar mass distribution of the clumps above our completeness limit inside a giant clump is consistent with random sampling of the CSMF. We will now build this result into a predictive framework -- the SCALES model -- in which the constituent clumps of a giant clump are simply random draws from the CSMF, summed until they reach the total mass of the giant clump. We emphasize that the SCALES model is a purely empirical and simplified model. It contains no physical prescription for how clumps form, evolve or interact, and is only a statistical description of the observed clump composition of giant clumps. We stress, however, that we do not interpret the draws below our completeness limit as clumps that are currently present as distinct structures, since these cannot be directly verified. Instead, they only make up the remaining mass budget of the giant clump, extrapolated from what we find for the clumps above $10^{7.1}\,\mathrm{M_\odot}$. These clumps might have formed, but not necessarily survived as distinct structures (e.g.\ due to mergers or disruption), with their mass still contained within the giant clump. It is worth noting that the model also completely ignores the effects of stellar feedback, which is known to affect the star formation at these scales by regulating the growth of the clumps and potentially disrupting them \citep{genel12, hopkins12, mandelker17, oklopcic17, fensch21}. 

With these caveats in mind, we do note that the model makes a specific, testable prediction: how dominant the most massive constituent clump should be, in terms of stellar mass, as a function of the mass of the host giant clump. In this section, we will use this prediction to understand the bridge between clumps and giant clumps empirically. In Fig.~\ref{fig:dominance_and_ratio} (left panel), we plot the stellar mass ratio of the most massive constituent clump to the host giant clump, as a function of the latter quantity for the 72 giant clumps in our sample. We then estimate the prediction of the purely empirical SCALES model by generating giant clumps through random draws from the CSMF until the mass of the giant clump is reached. Here, we assume the CSMF to follow a single power law (since the break may be an artifact of completeness; Sec.~\ref{subsec:mass_function}) with the high-mass slope of $1.88\pm0.04$ (as measured in Sec.~\ref{subsec:mass_function}). 

We limit the CSMF between $10^{9.5}\,\mathrm{M_\odot}$ and a lower cutoff ($M_{\rm min}$). Since the CSMF is not constrained below our completeness limit, we leave $M_{\rm min}$ as a free parameter, and vary it between $10^{3}$ and $10^{7}\,\mathrm{M_\odot}$ in steps of $0.25\,\mathrm{dex}$. We only retain the generated giant clumps that host at least one clump above $10^{7.1}\,\mathrm{M_\odot}$, following the criteria used in our study. Furthermore, we limit the number of giant clumps generated at each giant clump mass, including those that fail this selection, such that their total mass does not exceed the stellar mass of each of our 23 host galaxies (Table~\ref{tab:host_galaxies}), and repeat this over 20 realizations per galaxy. For each generated giant clump, the CSMF slope is randomly drawn within its uncertainty.

The model predicts the mass share of the most massive clump to decline with giant clump mass, approaching unity toward $10^{7.1}\,\mathrm{M_\odot}$. As observed in Fig.~\ref{fig:dominance_and_ratio}, this is mainly due to our selection however, which imposes $f_{\rm top1}\geq10^{7.1}\,\mathrm{M_\odot}/M_{\rm giant}$, and drives the predicted trend below $\approx10^{8}\,\mathrm{M_\odot}$ irrespective of $M_{\rm min}$. Meanwhile, toward higher giant clump masses, the predicted $f_{\rm top1}$ is instead set by $M_{\rm min}$. One can understand this as a lower cutoff requiring a larger number of constituent clumps to make up the giant clump mass. This reduces the mass share of the most massive constituent clump. 

We now quantify the agreement of our measured values with the SCALES model predictions with different $M_{\rm min}$ values using a Kolmogorov--Smirnov (KS) test. For each observed giant clump, we generate 300 model giant clumps of the same mass, filled with clumps drawn from the CSMF above $M_{\rm min}$ and subject to the same selection as the data. For each observed giant clump, we then count the fraction of its model predictions with a lower $f_{\rm top1}$ than the observed value, i.e.\ its percentile among the predictions. If the model describes the data, each observed giant clump is simply one more random realization of the model, and is therefore equally likely to fall anywhere among its predictions. Across our sample, the percentiles should then be evenly (uniformly) distributed between 0 and 1. A clustering toward 1 (or 0) would instead indicate that the observed giant clumps are systematically more (or less) dominated by their most massive clump than predicted. 

We test for uniformity using the KS test for each $M_{\rm min}$ (Fig.~\ref{fig:dominance_and_ratio}, right panel). To account for the measurement uncertainties, we resample the observed $f_{\rm top1}$ of each giant clump within its uncertainties and repeat the test 200 times. We quote the median $p$-value, and the 16th--84th percentile range of the $p$-values across the resamplings. We repeat this whole procedure also for the fraction of the giant clump mass held by the clumps above $10^{7.1}\,\mathrm{M_\odot}$ ($f_{\rm bright}$; the complement of the fraction shown in Fig.~\ref{fig:ec_budget}). This provides an additional metric to describe the composition of the giant clumps.

We find both quantities to reject a lower cutoff of $\leq10^{5}\,\mathrm{M_\odot}$, with a median $p\leq0.003$, and $p<10^{-3}$ for $\leq10^{4.25}\,\mathrm{M_\odot}$. Both metrics favor $M_{\rm min}\approx10^{6}\,\mathrm{M_\odot}$, with a median $p=0.10$ ($0.02$--$0.29$) and $0.04$ ($0.01$--$0.08$) for $f_{\rm top1}$ and $f_{\rm bright}$, respectively.\footnote{We note that the median $p$-value of $f_{\rm bright}$ for this cutoff lies slightly below $0.05$. However, its 16th--84th percentile range across the resamplings extends above $p=0.05$.} Hence, based on this analysis, we do not expect the CSMF of the clumps that make up a giant clump to continue with the measured high-mass slope much below $\approx10^{6}\,\mathrm{M_\odot}$, with an extension below $10^{5}\,\mathrm{M_\odot}$ being rejected. Instead, the CSMF likely flattens and cuts off around these masses. However, since these clumps might not survive as distinct structures (see above), this does not necessarily translate into the CSMF that would be observed with deeper observations.

It is worth noting that $M_{\rm min}$ does not necessarily mark the lowest mass of any star-forming structure inside a giant clump. In S-I, we argued that the clumps themselves are made up of smaller star-cluster-like substructures (Sec.~4.2.3 in S-I). Rather, $M_{\rm min}$ marks the lowest mass of the clumps that can form as constituents of a giant clump, although they might not survive as distinct structures over long timescales. This is analogous to the relation between the clumps and the giant clumps. The clumps are the constituents of the giant clumps and follow a consistent mass function slope (Sec.~\ref{subsec:mass_function}), but cannot be placed in the same mass function, since their stellar mass would then be counted twice. Similarly, the star clusters making up the clumps may follow the same slope, but form a separate population, with their stellar mass already included in that of the clumps. Hence, our results do not reject a scale-free hierarchy extending down to star cluster scales, but rather a single mass function in which the star clusters are drawn alongside the clumps as direct constituents of the giant clumps. This lower cutoff is also consistent with the mass budget in Sec.~\ref{subsec:mass_function}, where the EC mass is sufficient for the CSMF to continue with the high-mass slope down to $\approx10^{5.5}\,\mathrm{M_\odot}$ for the median values. For $M_{\rm min}\approx10^{6}\,\mathrm{M_\odot}$, the mass of the missing clumps ($\log(M_\star/\mathrm{M_\odot})\approx9.9$) lies slightly below that of the ECs ($\log(M_\star/\mathrm{M_\odot})=10.0\pm0.2$), with the remainder possibly contributed by the older clumps discussed in Sec.~\ref{subsec:mass_function}.

In conclusion, the mass composition of our giant clumps is consistent with the SCALES model, with the CSMF of their constituent clumps extending only down to $\approx10^{6}\,\mathrm{M_\odot}$. Without our selection, the model then predicts the median mass fraction of the most massive constituent clump to rise from $\approx0.2$ at a giant clump mass of $10^{9.5}\,\mathrm{M_\odot}$ to $\approx0.5$ at $10^{7.5}\,\mathrm{M_\odot}$ (dotted lines in Fig.~\ref{fig:dominance_and_ratio}), since fewer constituent clumps are required to make up the giant clump mass as it approaches $M_{\rm min}$. Hence, we predict low mass giant clumps to be more likely dominated by individual (lensed) clumps. 

We note that unlensed galaxy studies have concluded that the giant clump mass is a function of the total galaxy mass \citep{kalita25a,sok25}. Taking this as a purely empirical inference, we can thus claim that low mass galaxies, if observed with sufficient depths, can provide samples of low mass giant clumps that are likely to be dominated by single clumps. Hence, lensed studies targeting such low mass galaxies would find clumps with masses comparable to those of the giant clumps detected in unlensed studies. We find an example of this in \cite{kalita25a}, where in the lowest mass galaxies of their sample ($9.5<\log(M_{*}/\mathrm{M_\odot})<9.8$), the giant clumps span a mass range of $7.4<\log(M_{*}/\mathrm{M_\odot})<8.7$ -- values typically observed at the massive end of (lensed) clump mass distributions \citep[this work;][]{messa22, messa24, claeyssens23, claeyssens25}, even though the resolutions of these two groups of observations differ substantially. \cite{kalita25a} report a spatial detection limit of $0.5$--$1.2\,\mathrm{kpc}$ over their redshift range, compared to the $\approx50$--$400\,\mathrm{pc}$ sizes resolved for the (lensed) clumps in this work and in \cite{messa22, messa24, claeyssens23, claeyssens25}. This is a difference of at least $\sim100\,\mathrm{pc}$, and up to $\sim1\,\mathrm{kpc}$.

\begin{figure}
\centering
\includegraphics[width=0.47\textwidth]{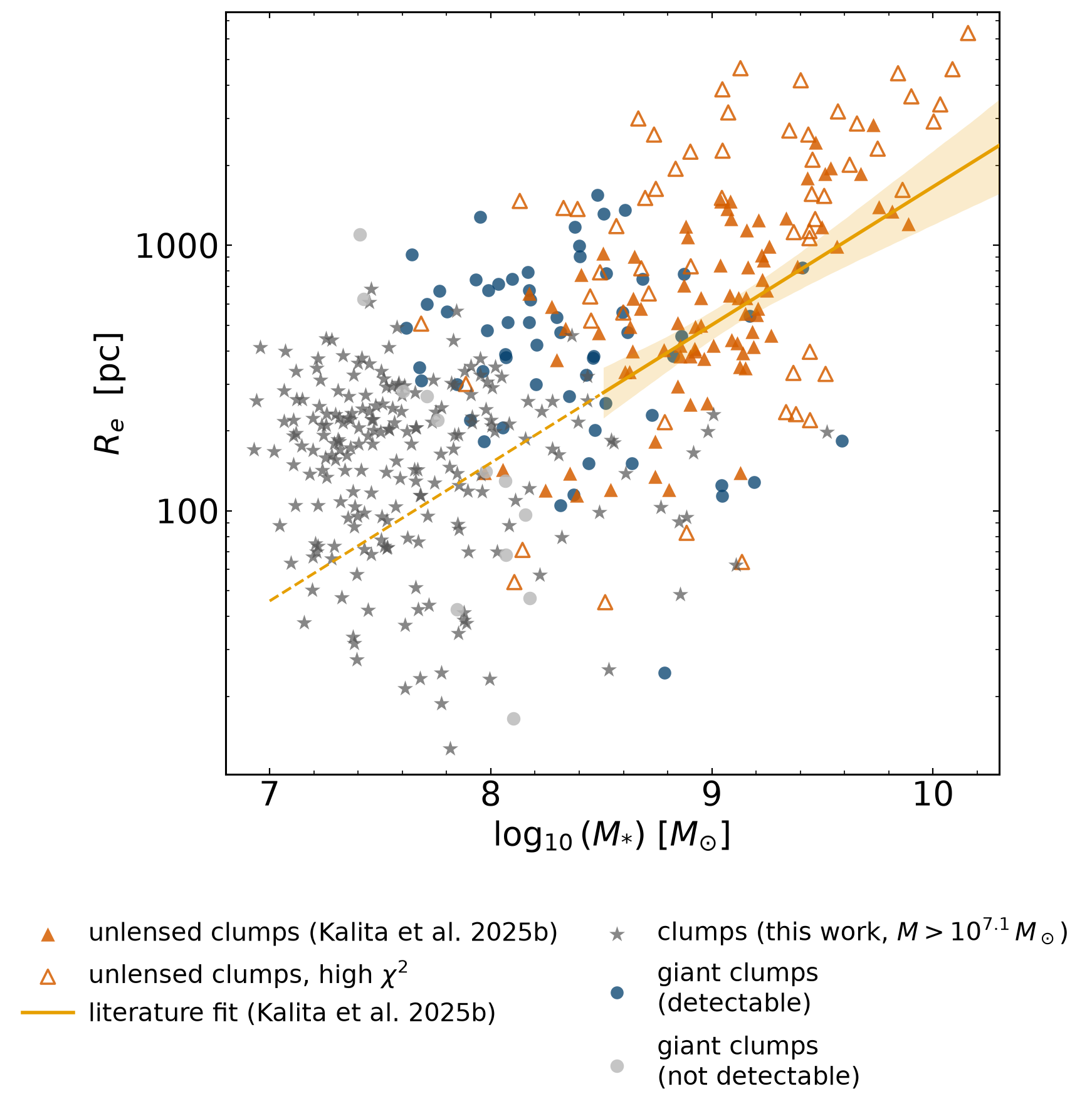}
\caption{The effective radius of our 72 giant clumps (Sec.~\ref{subsec:size_dominance_test}) as a function of total giant clump stellar mass, compared to the unlensed clump sample of \cite{kalita25b} (orange triangles; open triangles: their clumps with high reduced-$\chi^2$ SED fits) over the same mass range. Our giant clumps are split by whether they fall above (``detectable'', dark blue) or below (``not detectable'', light gray) the $90\%$ completeness limit of the \cite{kalita25b} sample. Gray stars show the individual bright ($M_\star>10^{7.1}\,\mathrm{M_\odot}$) lensed clumps in our own sample, each plotted at its own fitted mass and size (based on a single random draw within its uncertainties). The solid orange line is the \cite{kalita25b} power-law fit, extrapolated to lower masses as a dashed line, with the shaded region showing its $1\sigma$ uncertainty.}
\label{fig:mass_size}
\end{figure}

Hence, we deduce that one transitions from giant clumps to clumps as a result of the way clumps are organized within giant clumps. We illustrate this transition through a mass--size relation (Fig.~\ref{fig:mass_size}), for which we require the sizes of the giant clumps in our sample. Up to this point, we have not measured them, and have instead been referring to the sizes of the ECs, which are already systematically restricted to giant clump scales (Sec.~2.2 in S-I). Hence, for the purpose of this comparison, we estimate the sizes of the giant clumps, albeit without the rigor of the other measurements we have made. Using the best-fit 2D F070W models of the clumps and the EC that make up a giant clump, we generate a 2D flux distribution over a noiseless background. Centered on the resulting flux centroid, we fit an elliptical Gaussian with fixed ellipticity and phase angle (based on the lensing shear, similar to that used in the primary fitting; Sec.~\ref{subsec:det_flux_fit}). After correcting for the magnification at the location of the centroid, this gives us the effective radius of the giant clump. 

However, our rudimentary effort to `measure' the size of the giant clumps to compare to unlensed observations \citep[e.g.,][]{ji24, kalita25b} overlooks a key systematic aspect of such unlensed observations: the depth. Their detections are biased in favor of giant clumps that have high flux/mass surface densities. Meanwhile, in our analysis, we do not technically detect the giant clumps. We rather detect the ECs alongside clumps, and make the associations later. Although this means a perfect comparison cannot be made with the current analysis, we use the completeness limits as a function of total stellar mass and effective radius in \cite{kalita25b} (their detection-fraction map, where a giant clump above $90\%$ completeness will be considered detectable) to classify our giant clumps as `detectable' and `not detectable' in reference to their study. We plot these classified giant clumps in Fig.~\ref{fig:mass_size} in terms of their mass and size, and compare them to the unlensed clumps in \cite{kalita25b}. We caution that the two sets of giant/unlensed clumps come from galaxies with largely non-overlapping ranges of stellar mass --- their sample with mass $\sim 10^{10.5 - 11.5}\,\mathrm{M_\odot}$, against ours with values $\sim 10^{8.5 - 10.5}\,\mathrm{M_\odot}$ --- and as a result, likely form in very different ISM conditions. Hence, the comparison should be taken merely as qualitative.

With these caveats in mind, we find that the majority of our giant clumps (60/72) would be detectable in \cite{kalita25b}. Where the two samples overlap in mass ($\gtrsim10^{8.5}\,\mathrm{M_\odot}$), these giant clumps are broadly consistent with the unlensed clumps and their mass--size relation \citep{kalita25b}. Meanwhile, toward lower masses, they mostly lie above the extrapolation of this relation.  In order to illustrate the transition from giant clumps to clumps, we also place the $>10^{7.1}\,\mathrm{M_\odot}$ clumps from our sample. We find these to also be in agreement with the extrapolation of the same relation. However, the scatter of the giant clumps and the massive lensed clumps falls preferentially above the relation, toward lower mass surface densities. This is in line with our initial assessment of unlensed clump detections being biased toward higher surface densities, while our analysis reveals fainter structures. Furthermore, the mass--size relation for the unlensed clumps we show in the plot uses those with well-fit SEDs in \cite{kalita25b}. However, they also find additional structures they do not include in the main sample because their SED fits return high reduced-$\chi^2$ (shown with open triangles in Fig.~\ref{fig:mass_size}). Including these also increases the scatter of their sample above the relation (toward lower surface densities), in line with our results.

One key takeaway from Fig.~\ref{fig:mass_size}, as well as the rest of the giant clump composition analysis, is that the clumps and giant clumps are not distinct populations, but nested ones. They reflect a hierarchical arrangement of star-forming structures, similar to that observed in the local Universe, where star formation is organized hierarchically from individual clusters and associations up to $\sim\mathrm{kpc}$-scale star complexes \citep{efremov95, elmegreen96a, elmegreen96b, rodriguez20, larson23, shashank25}. We caution, however, that this local hierarchy is only approximately self-similar: its levels carry preferred scales set by distinct physical processes, with the largest scales dictated by the Toomre scale of their host disks \citep{elmegreen96a, elmegreen06, grasha17, menon21, turner22}. Our analysis, meanwhile, only establishes that the mass distribution of the clumps above our completeness limit inside giant clumps is consistent with random sampling of the CSMF (Sec.~\ref{subsec:stochastic_composition}), and the clumps themselves would likely be composed of even smaller star-cluster-like structures (S-I). This supports, but is not sufficient to establish, a hierarchy of physical structures as in the local Universe. For that, one would need to also resolve the sub-clump ($\approx 10\,\mathrm{pc}$) scales within the same giant clumps. We have also not established whether the largest scales are in fact physical (as in the local Universe), or simply random associations, as discussed in Sec.~\ref{subsec:giant_clump_models}. 

\subsection{Clumps as observational constraints on giant clump models}
\label{subsec:giant_clump_models}
In this section, we review the implications of a giant clump composition consistent with random sampling of the CSMF for the physical understanding of giant clumps. In the standard picture, giant clumps form in situ through violent disk instability (VDI): gas-rich, turbulent high-redshift disks fed by cold streams self-regulate to a Toomre parameter $Q\sim1$ and fragment into giant clumps.  Each giant clump is a few percent of the disk mass, with a characteristic Toomre or Jeans mass of order $10^{8}$--$10^{9}\,\mathrm{M_\odot}$ on $\sim\mathrm{kpc}$ scales \citep{dekel09, dekel14, bournaud14}. However, this picture has always carried an implicit statement about the constituent substructure (which we refer to as clumps in this work). \cite{dekel09} assumed the giant clump mass to be partitioned into equally massive (sub-)clumps, with this fragmented internal structure allowing feedback-driven winds to escape through the low-density channels. But the discussions on the constituent clumps of giant clumps were always limited to order-of-magnitude arguments, without specifying the expected mass distribution of these clumps.

The properties of the giant clumps constrain their internal content only indirectly.  The mass of a giant clump is set by the balance of accretion from the disk, star formation, feedback-driven outflows, and tidal stripping.  Within the framework where the giant clumps can survive for at least a few hundred Myr, \cite{dekel13} argue that the outflows are steady rather than disruptive. Growth driven predominantly by accretion onto the giant clump, along with (giant) clump -- (giant) clump mergers, means the structures grow by roughly a factor of a few during their lifetime \citep[see also][]{mandelker14}. By contrast, \cite{bournaud14} find that feedback-driven outflows scale steeply with the star-formation rate and roughly cancel out the growth, keeping the clump mass approximately constant throughout migration.  

However,  more important for our work,  both these groups conclude that there is an exchange of mass between giant clumps and their surroundings, with the more steady-state scenario of \cite{bournaud14} concluding that giant clumps act as overdense regions, similar to spiral arms in local galaxies, rather than a self-contained system. We argue that our observation of a global CSMF-based distribution of the clumps above our completeness limit inside giant clumps is consistent with this picture. This is further corroborated in \citet{faure21}, using parsec-scale simulations with matched mock observations. They find that the constituent clumps inside a giant clump are completely agnostic of the properties of the giant clump, qualitatively agreeing with our conclusions.

On the other hand,  \citet{oklopcic17}, using cosmological simulations with explicit radiative feedback, find that giant clumps are short-lived, disrupted within $\sim20\,\mathrm{Myr}$ \citep[see also][for isolated-disk simulations with similar feedback reaching the same conclusion]{hopkins12}.  However, this differing conclusion pertains only to the kpc-scale structure, without any claims of disagreement on the $\approx 50-400\,\mathrm{pc}$ clump scales within. In fact, these constituent clumps are instead highlighted as playing a more prominent role in determining the dynamics of the host giant clump, with the latter being described as ``loose complexes of bound (sub-)clumps'' \citep[see also][]{behrendt16}.

It is worth noting here that the giant clumps being an emergent feature also allows for a connection to the hierarchical organization of star formation in the local Universe. \citet{tamburello17} show that disk fragmentation produces clumps with an intrinsic characteristic mass and size similar to those of local giant molecular-cloud complexes. In these more `clump-centric' frameworks, the CSMF-based composition of giant clumps fits naturally, since they are usually regarded as associations without a dominant physical significance.

Hence, although simulations disagree on the giant clump scales, the predictions regarding the constituent clumps within have been mainly limited to order-of-magnitude statements about their stellar mass and size. As a result, it would be unjustified to make any claims regarding which scenario our observational result supports. Furthermore, although the current work discussed the distribution of mass across the clumps inside giant clumps, the parallel effort of quantifying their spatial distribution will be carried out in the next work of the SCALES series (Kalita et al. in prep.). It is also worth noting that our analysis, including the purely empirical SCALES model, does not account for stellar feedback, even though it is central to these models, both through the outflows that regulate the growth of giant clumps \citep{dekel13, bournaud14} and through the disruption that sets their lifetimes \citep{genel12, hopkins12, oklopcic17}. Our results therefore cannot discriminate between these scenarios on the basis of feedback. Hence, we only provide an observational constraint on the mass composition of giant clumps, against which the various aforementioned models can be re-evaluated.

\subsection{Impact of the possible bulge candidates}
\label{subsec:bulge_impact}

As in S-I, we do not remove the clumps that may correspond to a central bulge, since such an identification depends on the galaxy in the image plane resembling a disk--bulge configuration. In S-I, we visually identify 7 such clumps in galaxies 7775, 15439, 20054 and 22864, four of which are the most massive clumps in their hosts. These are associated with 4 of our 72 giant clumps. Here, we assess the impact of removing these possible bulge candidates on our results.

Excluding them steepens the CSMF slope to $1.94\pm0.05$ (with $M_b = 10^{7.53\pm0.05}\,\mathrm{M_\odot}$; S-I), and reduces our sample to 70 giant clumps, since two of the giant clumps lose their only clump above $10^{7.1}\,\mathrm{M_\odot}$. The composition of the giant clumps however remains effectively unchanged. A single clump holds the majority of the giant clump mass in $27\%$ of the giant clumps (compared to $29\%$ with the bulge candidates), and the two most massive clumps in $54\%$ of those hosting at least two clumps above $10^{7.1}\,\mathrm{M_\odot}$ (compared to $56\%$). The combination of the EC and the faint clumps dominates in $41\%$ of the giant clumps (compared to $43\%$), dropping to $24\%$ (compared to $25\%$) for the highly conservative lower limit. The stratified test also finds the observed configuration of the clumps above $10^{7.1}\,\mathrm{M_\odot}$ to be consistent with any random reassignment of their stellar masses ($p = 0.71$). Meanwhile, the giant clump mass function follows a slope of $1.94\pm0.11$, effectively unchanged from the $1.95\pm0.15$ with the bulge candidates.

The mass budget argument (Sec.~\ref{subsec:mass_function}) is the most affected. Since the steeper CSMF requires more mass below the break to sustain the high-mass slope, the median EC mass fills the deficit only for cutoffs above $\approx10^{6}\,\mathrm{M_\odot}$ (compared to $\approx10^{5.5}\,\mathrm{M_\odot}$), while the two overlap within the uncertainties down to $\approx10^{4.5}\,\mathrm{M_\odot}$, rather than $10^{3}\,\mathrm{M_\odot}$. Nevertheless, the EC mass remains sufficient for the CSMF to continue with the high-mass slope well below the break. Similarly, repeating the comparison with the SCALES model (Sec.~\ref{subsec:size_dominance_test}) with the steeper CSMF slope shifts the lower cutoff that agrees best with our giant clumps slightly upwards, to $M_{\rm min}\approx10^{6.25}\,\mathrm{M_\odot}$. For this cutoff, we find a median $p=0.10$ ($0.03$--$0.30$) and $0.04$ ($0.02$--$0.10$) for $f_{\rm top1}$ and $f_{\rm bright}$, respectively, comparable to those for $M_{\rm min}\approx10^{6}\,\mathrm{M_\odot}$ with the bulge candidates. Meanwhile, both quantities reject a lower cutoff of $\leq10^{5}\,\mathrm{M_\odot}$ even more strongly, with a median $p<10^{-3}$. Hence, removing the possible bulge candidates does not affect the conclusions of this work.

\section{Summary}
\label{sec:summary}

This paper (SCALES-II) asks what the internal substructure of giant star-forming clumps at $z>1$ is.  Using 23 lensed galaxies at $z=1-4$, imaged with JWST/NIRCam,  we implement a newly developed difference-of-Gaussian decomposition that simultaneously models the compact clumps ($\approx 50-400\,\mathrm{pc}$) and the extended component at giant clump scales (the ECs) in which they sit (method presented in detail in S-I). In this work, we associate clumps with ECs by spatial proximity to reconstruct 72 giant clumps and decompose their stellar mass among the constituents. The main results and conclusions are as follows:

\begin{enumerate}
    \item A single clump holds the majority of the mass in roughly $29\%$ of the giant clumps, while in $56\%$ of those hosting at least two clumps above $10^{7.1}\,\mathrm{M_\odot}$, the two most massive constituent clumps make up the majority. 
    \item The CSMF breaks from a slope of $1.88$ to a shallower slope below $\sim10^{7.5}\,\mathrm{M_\odot}$, close to our completeness limit. Accounting for the mass budget of the ECs, the break may be an artifact of clumps blended into the ECs. Within the uncertainties, the EC mass is sufficient for the high-mass slope to continue below the break down to a lower cutoff of $\sim10^{3}\,\mathrm{M_\odot}$, although the median values require a cutoff above $\sim10^{5.5}\,\mathrm{M_\odot}$. Furthermore, the giant clumps follow a slope ($1.95\pm0.15$) consistent with that of the CSMF, giving the same scale-free mass function slope at (lensed) clump and giant clump scales.
    \item A within-galaxy mass-shuffling test based on the most massive clump within a giant clump shows the composition of giant clumps, as traced by their clumps above our completeness limit ($10^{7.1}\,\mathrm{M_\odot}$), to be statistically consistent with random sampling of the CSMF.  Hence, these constituent clumps can be generated by a simple random draw from the full population of clumps in a galaxy, and hold no information about being part of a giant clump. However, whether this extends to the clumps below the completeness limit remains uncertain, and the large measurement uncertainties might hide additional internal effects.
    \item Based on the previous result, we create a simple statistical model in which the constituent clumps of a giant clump are randomly drawn from the CSMF until their summed mass reaches the giant clump mass. We find this model to be consistent with the mass composition of our giant clumps, provided that the CSMF of the constituent clumps extends only down to $\approx10^{6}\,\mathrm{M_\odot}$, with an extension below $10^{5}\,\mathrm{M_\odot}$ being rejected. Hence, we predict that even smaller structures (possibly star clusters) are not direct constituents of the giant clumps, and more likely form a separate population as constituents of the clumps that in turn make up the giant clumps. 
    \item We find that the CSMF-sampled composition of giant clumps is qualitatively consistent with both a bound giant clump scenario with constant renewal of its constituents and giant clumps being an emergent feature due to the spatial distribution of the constituent clumps, but cannot discriminate between them. This is mainly due to the relevant models being ambiguous in their predictions of the mass distribution among the substructure of giant clumps beyond order-of-magnitude statements about their mass and size. Hence, we propose our observations to be treated as constraints on future analyses of such models.

\end{enumerate}
\begin{acknowledgments}
We would like to thank Lilan Yang, Oleg Gnedin, Reinhard Genzel and Giulia Tozzi for the highly valuable discussions. BSK was supported by the National Science Foundation of China (W2633004), the China Postdoctoral Science Foundation (2025M773186), and the Beijing Natural Science Foundation (IS26015). LCH was supported by the National Science Foundation of China (12233001) and the China Manned Space Program (CMS-CSST-2025-A09). Finally, BSK would like to thank Xueling Wang for her constant support, without whom this work would not have been possible.

\textit{Disclosure of Delegation to Generative AI:} The authors declare the use of generative AI in the research and writing process. According to the GAIDeT taxonomy (2026), the following tasks were delegated to GAI tools under full human supervision:
\begin{itemize}[noitemsep]
    \item Code generation
    \item Code optimization
    \item Process automation
    \item Proofreading and editing
\end{itemize}
The GAI tool used was: Claude Sonnet. Responsibility for the final manuscript lies entirely with the authors. GAI tools are not listed as authors and do not bear responsibility for the final outcomes. Declaration submitted by: Boris S. Kalita.
\end{acknowledgments}

\begin{contribution}
BSK carried out the analysis and prepared the manuscript. AC provided supporting datasets from a separate analysis of the parent sample. AC, LCH, MDZ, NMFS and ED provided key insights during the planning phase of the project. All co-authors provided valuable comments during the preparation of the manuscript.
\end{contribution}

\appendix

\section{EC-Clump Associations}
\label{app:ec_associations}
\begin{figure*}[!htp]
\centering
\includegraphics[width=\textwidth]{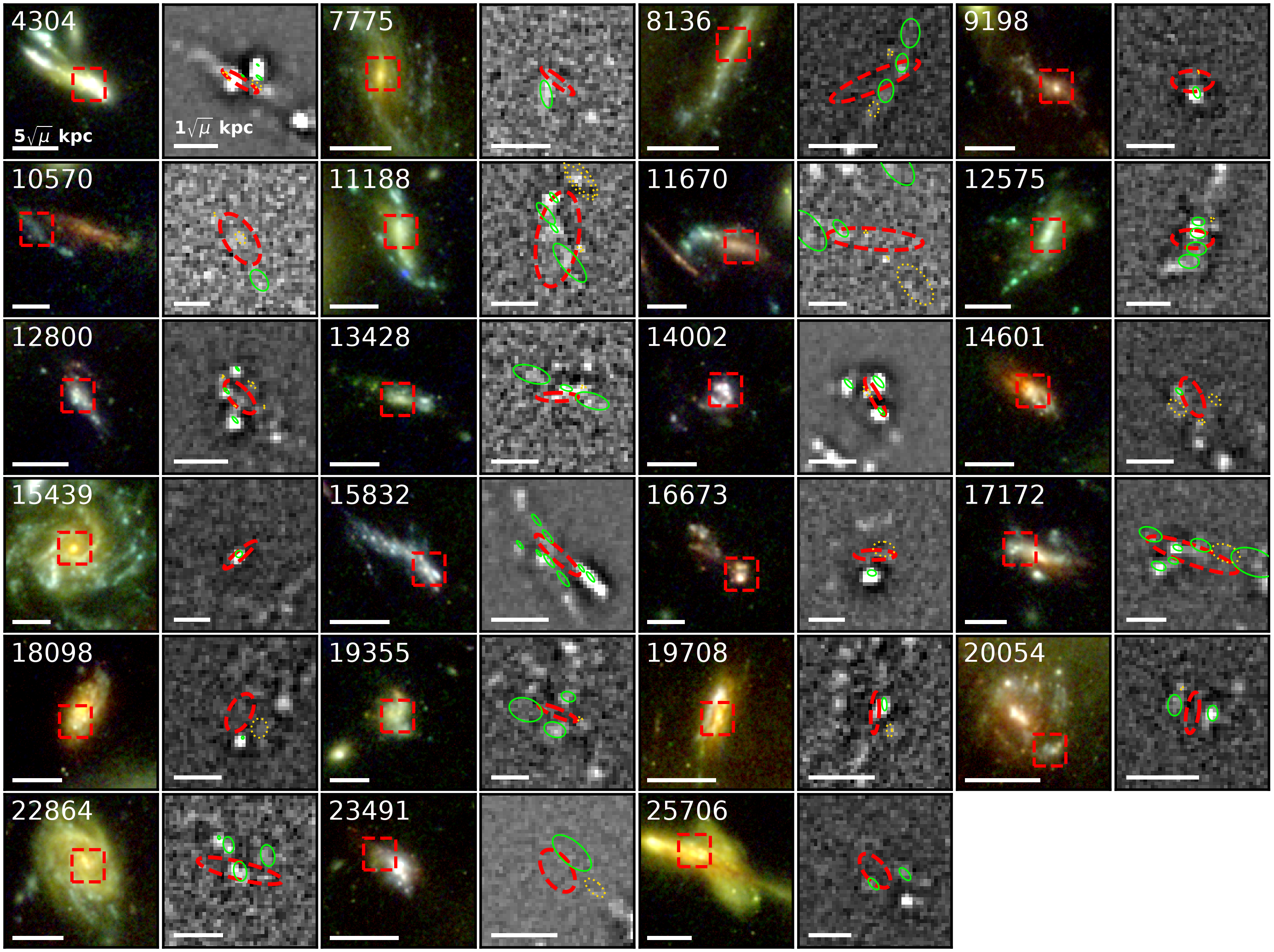}
\caption{A collage showing examples of EC-clump associations across multiple galaxies in the sample. The arrangement is similar to that in Fig.~\ref{fig:galaxy_illustration}. The left panel shows the RGB (F200W/F115W/F070W) image of the galaxy, with one of its giant clumps being highlighted with a red box. The F070W image region shown in the box is then shown in the right panel. This image is generated by adding the 5 smallest DoG planes. In this panel, the EC is shown in red, while the green ellipses show the location of the associated clumps (dotted gold: associated clumps without an SED fit, whose light is fitted with the EC). Collectively, they make up a giant clump.}
\label{fig:ec_assoc_collage}
\end{figure*}

Here in Fig.~\ref{fig:ec_assoc_collage}, we provide examples of giant clumps in all 23 galaxies in our sample.

\bibliography{main}{}
\bibliographystyle{aasjournalv7}

\end{document}